\documentclass[11pt,a4paper]{article}
\pdfoutput=1
\usepackage{jheppub}
\usepackage{amsfonts}
\usepackage{amsmath}
\usepackage{array}
\usepackage[markup=nocolor]{changes}
\usepackage{graphicx}
\usepackage{hyperref}
\usepackage{multirow}
\usepackage{url}
\usepackage[table]{xcolor}
\usepackage{xspace}

\newcolumntype{C}[1]{>{\centering\let\newline\\\arraybackslash\hspace{0pt}}m{#1}}

\newcommand*\oline[1]{
  \hspace*{0.2em}
  \vbox{
    \kern-0.35ex
    \hrule height 0.4pt
    \kern0.35ex
    \hbox{
      \kern-0.5em
      \ifmmode#1\else\ensuremath{#1}\fi
      \kern-0.0em
}}}
\newcommand{\sigbr}{$\sigma \cdot {\cal B}$\xspace}
\newcommand{\HEPfit}{\texttt{HEPfit}\xspace}

\hypersetup{pdfstartview=FitV,colorlinks=true,linkcolor=blue,citecolor=red,filecolor=black,urlcolor=blue}

\definecolor{color1}{HTML}{cc9966}
\definecolor{color2}{HTML}{ff0000}
\definecolor{color3}{HTML}{dddd00}
\definecolor{color4}{HTML}{33ff00}
\definecolor{color5}{HTML}{ff00dd}
\definecolor{color6}{HTML}{0044dd}
\definecolor{color7}{HTML}{8822cc}
\definecolor{color8}{HTML}{00ffff}
\definecolor{color9}{HTML}{aa5511}
\definecolor{color10}{HTML}{9999cc}
\definecolor{color11}{HTML}{225522}
\definecolor{color12}{HTML}{ffa322}
\definecolor{sigstrcolor1}{HTML}{ee0000}
\definecolor{sigstrcolor2}{HTML}{00aaaa}
\definecolor{sigstrcolor3}{HTML}{550099}
\definecolor{sigstrcolor4}{HTML}{2255ff}
\definecolor{sigstrcolor5}{HTML}{22cc00}
\definecolor{sigstrcolor6}{HTML}{ff9900}

\title{Update of global Two-Higgs-Doublet model fits}

\author[a,b,c]{Debtosh Chowdhury,}
\emailAdd{Debtosh.Chowdhury@polytechnique.edu}

\author[d]{Otto Eberhardt}
\emailAdd{otto.eberhardt@ific.uv.es}

\affiliation[a]{Istituto Nazionale di Fisica Nucleare,
Sezione di Roma, Piazzale Aldo Moro 2, I-00185 Roma, Italy}

\affiliation[b]{Centre de Physique Th{\'e}orique, {\'E}cole Polytechnique, F-91128 Palaiseau Cedex, France}

\affiliation[c]{Laboratoire de Physique Th{\'e}orique, Universit{\'e} Paris-Sud, F-91405 Orsay Cedex, France}

\affiliation[d]{Instituto de F\'isica Corpuscular, Parque Cient\'ifico, 
C/Catedr\'atico Jos\'e Beltr\'an, 2, E-46980 Paterna, Spain}

\abstract{
We perform global fits of Two-Higgs-Doublet models with a softly broken $\mathbb{Z}_2$ symmetry to recent results from the LHC detectors CMS and ATLAS, that is signal strengths and direct search limits obtained at $\sqrt{s} = 8$ TeV and $\sqrt{s}=13$ TeV. We combine all available ATLAS and CMS constraints with the other relevant theoretical and experimental bounds and present the latest limits on the model parameters. We obtain that deviations from the so-called alignment limit $\beta-\alpha=\pi/2$ cannot be larger than $0.03$ in type I and have to be smaller than $0.02$ in the remaining three types. For the latter, we also observe lower limits on the heavy Higgs masses in the global fit. The splittings between these masses cannot exceed $200$ GeV in the types I and X and $130$ GeV in the types II and Y. Finally, we find that the decay widths of the heavy Higgs particles cannot be larger than $7\%$ of their masses if they are lighter than $1.5$ TeV.}

\preprint{{\raggedleft CPHT-RR084.112017 \\ LPT-Orsay-17-48 \\ IFIC/17-49 \par}}

\begin{document}
\maketitle

\section{Introduction}
\label{sec:intro}

The discovery of a new scalar resonance with a mass around 125 GeV \cite{Aad:2012tfa,Chatrchyan:2012xdj} in the Run 1 phase of the Large Hadron Collider (LHC) has paved the way for new directions in high-energy particle physics. Analyzing the properties of this particle has suggested strong evidence that it is the Higgs boson of the Standard Model (SM), i.e.~a scalar CP-even state which has SM-like couplings to the other particles. Currently the combined analysis based on the Run 1 (7 and 8 TeV) LHC data  shows that its couplings with the vector bosons are found to be compatible with those expected from the SM within a $\sim \! 10\%$ uncertainty, whereas the coupling to the third generation fermions (top, bottom quarks and the $\tau$ lepton) is compatible within an uncertainty of $\sim \! 15-20\%$ \cite{Khachatryan:2016vau}. Thus the current status of the Higgs properties still allows to explore new interpretations of the observation coming from new physics of different underlying structures.  

The Two-Higgs-doublet model (2HDM) \cite{Lee:1973iz,Gunion:2002zf,Branco:2011iw} is one of such extensions of the Standard Model. Alike other popular NP models, it gets more and more constrained by recent experimental progress, especially by the LHC data \cite{Chen:2013kt,Chiang:2013ixa,Grinstein:2013npa,Barroso:2013zxa,Coleppa:2013dya,Eberhardt:2013uba,Belanger:2013xza,Chang:2013ona,Cheung:2013rva,Celis:2013ixa,Wang:2013sha,Baglio:2014nea,Inoue:2014nva,Dumont:2014wha,Kanemura:2014bqa,Ferreira:2014sld,Broggio:2014mna,Dumont:2014kna,Bernon:2014nxa,Chen:2015gaa,Chowdhury:2015yja,Craig:2015jba,Bernon:2015qea,Bernon:2015wef,Cacchio:2016qyh,Belusca-Maito:2016dqe}. As the name suggests the 2HDM has two Higgs doublets in contrast to the single Higgs doublet in the SM. This extension of the Higgs sector leads to the existence of five scalar bosons, namely a heavy and light CP-even Higgs boson, $H$ and $h$, a CP-odd Higgs boson, $A$, and a pair of charged Higgs bosons, $H^{\pm}$. Whether the scalar boson observed in the Run 1 of LHC is a part of an extended Higgs sector is an outstanding question and is at the cynosure of attention of the current Run 2 (13 TeV) phase of the LHC.

The questions we ask is: Which parts of the 2HDM parameter space are favoured after imposing the latest experimental data from the LHC?
Compared to the Run 1 phase of the LHC, in Run 2 the situation has changed in several respects: 
Alongside the latest Higgs signal strength data by the ATLAS and CMS collaborations also many of the recent results of searches for additional heavy Higgs bosons are more constraining than the Run 1 data. Both experiments have performed dedicated searches for new signatures in various possible final states at the LHC: besides the fermionic final states $t\bar t$, $b\bar b$, $\tau^+\tau^-$, $t\bar b$ and $\tau^+ \nu$ they include gauge bosons ($\gamma \gamma$, $Z \gamma$, $ZZ$, $W^+W^-$) and Higgs particles ($hh$, $hZ$, $HZ$, $AZ$) as the decay products of a heavy resonance.
So far, these searches for heavy resonances have remained elusive in the ATLAS and CMS data, and thus the measurements put model-independent $95\%$ C.L.~upper limits on the production cross section times branching ratios for different production processes and decay modes. In the present work, we assess the status of all four types of softly broken $\mathbb{Z}_2$ symmetric 2HDM when all the experimental constraints coming from the latest LHC data are taken into account. We confront these with the theoretical constraints on these models (positivity, stability and next-to-leading order unitarity). Furthermore, we perform global Bayesian fits to all relevant constraints on these models, which also include electroweak precision and flavour observables, and highlight the complementarity between them.

This paper is organized as follows: 
The 2HDM is defined in Section \ref{sec:model}. In Section \ref{sec:constraints} we list all relevant constraints and explain the fitting set-up. The results are presented in the subsequent sections, first taking into account only the Higgs signal strengths in Section \ref{sec:signalstrengths} and the direct searches in Section \ref{sec:heavy}, before combining them with the other constraints in Section \ref{sec:allconstraints}. We conclude in Section \ref{sec:conclusions}. In Appendix \ref{sec:appendixA} we explain how we treat the prior dependence of the massive parameters.

\section{Model}
\label{sec:model}

The Two-Higgs-Doublet model with a softly broken $\mathbb{Z}_2$ symmetry is characterized by the following scalar potential:
\begin{align}
 V
 &=m_{11}^2\Phi_1^\dagger\Phi_1^{\phantom{\dagger}}
   +m_{22}^2\Phi_2^\dagger\Phi_2^{\phantom{\dagger}}
   -m_{12}^2 ( \Phi_1^\dagger\Phi_2^{\phantom{\dagger}}
              +\Phi_2^\dagger\Phi_1^{\phantom{\dagger}})
   +\tfrac12 \lambda_1(\Phi_1^\dagger\Phi_1^{\phantom{\dagger}})^2
   +\tfrac12 \lambda_2(\Phi_2^\dagger\Phi_2^{\phantom{\dagger}})^2
 \nonumber \\
 &\phantom{{}={}}
  +\lambda_3(\Phi_1^\dagger\Phi_1^{\phantom{\dagger}})
            (\Phi_2^\dagger\Phi_2^{\phantom{\dagger}})
  +\lambda_4(\Phi_1^\dagger\Phi_2^{\phantom{\dagger}})
            (\Phi_2^\dagger\Phi_1^{\phantom{\dagger}})
  +\tfrac12 \lambda_5 \left[ (\Phi_1^\dagger\Phi_2^{\phantom{\dagger}})^2
                      +(\Phi_2^\dagger\Phi_1^{\phantom{\dagger}})^2 \right],\label{eq:pot}
\end{align}
where $\Phi_1$ and $\Phi_2$ are the two Higgs doublets. While writing the potential we have assumed that the scalar potential is CP conserving. Instead of the eight potential parameters from Eq.~\eqref{eq:pot} we will use the physical parameters in the rest of this article. They consist of the vacuum expectation value $v$, the CP-even Higgs masses $m_h$ and $m_H$, the CP-odd Higgs mass $m_A$, the mass of the charged Higgs, $m_{H^+}$, the two diagonalization angles $\alpha$ and $\beta$, and the soft $\mathbb{Z}_2$ breaking parameter $m_{12}^2$. Assuming the observed scalar of mass $\sim \! 125$ GeV at the LHC to be the lighter CP-even Higgs $h$, the first two of these can be treated as fixed. The rest of the scalar masses could in general even be lighter than 125 GeV, they are not necessarily in the decoupling limit \cite{Gunion:2002zf}. Keeping in mind the discovery potential of the HL-LHC, in the following we will consider them to be in the range between 130 GeV and 1.6 TeV, that is beyond the region where the 125 GeV scalar was found. Moreover, we trade the angles $\alpha$ and $\beta$ with $\beta-\alpha$ and $\tan \beta$, since these combinations can be directly related to physical observables. All SM parameters were fixed to their best-fit values \cite{deBlas:2016ojx,deBlas:2017wmn}.

Neglecting the first two generations of fermions, the Yukawa part of the 2HDM Lagrangian reads as follows:
\begin{align*}
 {\cal L}_Y =& -Y_t \oline Q_{\textit{\tiny{L}}} i\sigma_2 \Phi_2^* t_{\textit{\tiny{R}}} -Y_{b,1} \oline Q_{\textit{\tiny{L}}} \Phi_1 b_{\textit{\tiny{R}}} -Y_{b,2} \oline Q_{\textit{\tiny{L}}} \Phi_2 b_{\textit{\tiny{R}}} -Y_{\tau,1} \oline L_{\textit{\tiny{L}}} \Phi_1 \tau_{\textit{\tiny{R}}} -Y_{\tau,2} \oline L_{\textit{\tiny{L}}} \Phi_2 \tau_{\textit{\tiny{R}}} + {\text{h.c.}}
\end{align*}
In the above Lagrangian, by convention the top quark only couples to $\Phi_2$; its Yukawa coupling is related to the SM value $Y_t^{\rm SM}$ by $Y_t\!=\!Y_t^{\rm SM}/\sin \beta$. With an unbroken $\mathbb{Z}_2$ symmetry in the Yukawa sector, there are only four possibilities through which the Higgs fields couple to the bottom quark and tau lepton at tree-level. They are called type I, type II, type X or ``lepton specific'' and type Y or ``flipped''. In Table \ref{tab:types} we categorize the corresponding Yukawa coupling assignments.
\begin{table}[htb]
  \centering
  \caption{Yukawa coupling assignments in the four possible $\mathbb{Z}_2$ symmetric 2HDM types.}\vspace{0.2cm}
  \begin{tabular}{l|l|l|l}
    \hline\hline
      Type I & Type II & Type X (``lepton specific'') & Type Y (``flipped'') \\
    \hline
      $Y_{b,1}=Y_{\tau,1}=0$ & $Y_{b,2}=Y_{\tau,2}=0$ & $Y_{b,1}=Y_{\tau,2}=0$ & $Y_{b,2}=Y_{\tau,1}=0$ \\
      $Y_{b,2}\!=\!Y_b^{\rm SM}/\sin \beta$ & $Y_{b,1}\!=\!Y_b^{\rm SM}/\cos \beta$ & $Y_{b,2}\!=\!Y_b^{\rm SM}/\sin \beta$ & $Y_{b,1}\!=\!Y_b^{\rm SM}/\cos \beta$ \\
      $Y_{\tau,2}\!=\!Y_\tau^{\rm SM}/\sin \beta$ & $Y_{\tau,1}\!=\!Y_\tau^{\rm SM}/\cos \beta$ & $Y_{\tau,1}\!=\!Y_\tau^{\rm SM}/\cos \beta$ & $Y_{\tau,2}\!=\!Y_\tau^{\rm SM}/\sin \beta$ \\
      \hline \hline
  \end{tabular}
  \label{tab:types}
\end{table}

\section{Constraints and fitting set-up}
\label{sec:constraints}

Our statistical analysis of the 2HDM is a Bayesian fit, in which the following priors are used for the previously defined parameters:
\begin{align*}
-1.1 & \leq \log(\tan \beta) \leq 1.7 \quad (\text{equivalent to } 0.08 \leq \tan \beta \leq 50),\\
0 & \leq \beta-\alpha \leq \pi,\\
130 \text{ GeV} & \leq m_H,m_A,m_{H^+} \leq 1.6 \text{ TeV},\\
-(1.6 \text{ TeV})^2 & \leq m_{12}^2 \leq (1.6 \text{ TeV})^2\\
\end{align*}
Extreme $\tan \beta$ values outside the chosen prior are expected to be excluded due to the absence of strong 2HDM effects in certain flavour observables (see e.g.~reference \cite{Deschamps:2009rh}); the aforementioned interval is a very conservative estimate. The only implicit assumption we make is that the $125$ GeV scalar is the \textit{light} CP-even Higgs particle of the 2HDM and that the other scalars should be heavier, yet in LHC reach.\\
The focus of this article is on LHC Higgs observables, that is $h$ signal strengths and searches for $H$, $A$ and $H^+$. Most details of the implementation of the corresponding observables can be found in our last article \cite{Cacchio:2016qyh}. The modifications to this will be explained in the following.

\begin{table}
\begin{center}
\begin{tabular}{| l | c | c | cccccc |}
\hline
& \textbf{Signal} & \textbf{Value} & \multicolumn{6}{c|}{\textbf{Correlation matrix}}\\[3pt]
& \textbf{strength} &  &&&&&&\\[3pt]
\hline
\hline
\cellcolor{sigstrcolor1} & $\mu_\text{ggF}^{\gamma \gamma}$ & $1.10\pm 0.23$  &1&-0.25&0&-0.14&0&\\[3pt]
\cellcolor{sigstrcolor1} & $\mu_\text{VBF}^{\gamma \gamma}$ & $1.3\pm 0.5$  &-0.25&1&0&0&0&\\[3pt]
\cellcolor{sigstrcolor1} & $\mu_\text{Wh}^{\gamma \gamma}$ & $0.5\pm 1.3$  &0&0&1&-0.64&0&\\[3pt]
\cellcolor{sigstrcolor1} & $\mu_\text{Zh}^{\gamma \gamma}$ & $0.5\pm 2.8$  &-0.14&0&-0.64&1&-0.11&\\[3pt]
\cellcolor{sigstrcolor1} & $\mu_\text{tth}^{\gamma \gamma}$ & $2.2\pm 1.5$  &0&0&0&-0.11&1&\\[3pt]
\hline
\hline
\cellcolor{sigstrcolor5} & $\mu_\text{ggF}^{ZZ}$ & $1.13\pm 0.33$  &1&-0.26&&&&\\[3pt]
\cellcolor{sigstrcolor5} & $\mu_\text{VBF}^{ZZ}$ & $0.1\pm 0.9$  &-0.26&1&&&&\\[3pt]
\hline
\hline
\cellcolor{sigstrcolor4} & $\mu_\text{ggF}^{WW}$ & $0.84\pm 0.17$  &1&-0.16&&&&\\[3pt]
\cellcolor{sigstrcolor4} & $\mu_\text{VBF}^{WW}$ & $1.2\pm 0.4$  &-0.16&1&&&&\\[3pt]
\hline
\cellcolor{sigstrcolor3} & $\mu_\text{ggF}^{\tau \tau}$ & $1.0\pm 0.6$  &1&-0.37&0&-0.25&0&-0.21\\[3pt]
\cellcolor{sigstrcolor3} & $\mu_\text{VBF}^{\tau \tau}$ & $1.3\pm 0.4$  &-0.37&1&0&0&0&0\\[3pt]
\cellcolor{sigstrcolor4} & $\mu_\text{Wh}^{WW}$ & $1.6\pm 1.1$  &0&0&1&-0.12&-0.12&0\\[3pt]
\cellcolor{sigstrcolor3} & $\mu_\text{Wh}^{\tau \tau}$ & $-1.4\pm 1.4$  &-0.25&0&-0.12&1&0&0\\[3pt]
\cellcolor{sigstrcolor4} & $\mu_\text{Zh}^{WW}$ & $5.9\pm 2.4$  &0&0&-0.12&0&1&0\\[3pt]
\cellcolor{sigstrcolor3} & $\mu_\text{Zh}^{\tau \tau}$ & $2.2\pm 2.0$  &-0.21&0&0&0&0&1\\[3pt]
\hline
\cellcolor{sigstrcolor4} & $\mu_\text{tth}^{WW}$ & $5.0\pm 1.8$  &1&-0.47&&&&\\[3pt]
\cellcolor{sigstrcolor3} & $\mu_\text{tth}^{\tau \tau}$ & $-1.9\pm 3.5$  &-0.47&1&&&&\\[3pt]
\hline
\hline
\cellcolor{sigstrcolor2} & $\mu_\text{Wh}^{bb}$ & $1.0\pm 0.5$ &&&&&&\\[3pt]
\hline
\cellcolor{sigstrcolor2} & $\mu_\text{Zh}^{bb}$ & $0.4\pm 0.4$ &&&&&&\\[3pt]
\hline
\cellcolor{sigstrcolor2} & $\mu_\text{tth}^{bb}$ & $1.1\pm 1.0$ &&&&&&\\[3pt]
\hline
\hline
\cellcolor{sigstrcolor6} & $\mu_\text{pp}^{\mu \mu}$ & $0.1\pm 2.5$ &&&&&&\\[3pt]
\hline
\end{tabular}
\caption{$h$ signal strengths from Table 8, Table 13 and Figure 27 of the official ATLAS and CMS combination for Run 1 \cite{Khachatryan:2016vau}, based on 25 fb$^{-1}$ of integrated luminosity. We neglect correlations below 0.1. The colours in the first column indicate the decay category in Figures \ref{fig:SigStr} and  \ref{fig:rgagavsrg}.}
\label{tab:signalstrengths8}
\end{center}
\end{table}

For the signal strengths, we define $\mu _{\rm production}^{\rm decay}$, where ``${\rm production}$'' stands for the ggF, VBF, Vh, Zh, Wh, tth or pp production channels of the $h$, while ``${\rm decay}$'' denotes the subsequent $h$ decay products $\gamma \gamma$, $ZZ$, $WW$, $\tau\tau$, $bb$, $\mu\mu$ or $Z\gamma$.\footnote{In order to improve readability, we drop charge or conjugation labels when there is no ambiguity.} For the last one, only upper limits are available; we assign to this signal strength a central value of 0 and adjust the Gaussian error such that the likelihood distribution has the 95\% limit at the value provided by the experimental collaborations.
All $h$ couplings are calculated at leading order: While the fermionic decays and the bosonic decays to $WW$ and $ZZ$ are possible at tree-level, we apply one-loop expressions for the decays into final states including massless bosons (that is $gg$, $\gamma \gamma$ and $Z\gamma$) \cite{Gunion:1989we}.

\begin{table}
\begin{center}
\begin{tabular}{| l | c | c | cccc | c | c |}
\hline
& \textbf{Signal} & \textbf{Value} & \multicolumn{4}{c|}{\textbf{Correlation matrix}} & ${\cal L}$ & \textbf{Source}\\[3pt]
& \textbf{strength} &  &&&&&\textbf{[fb$^{-1}$]}&\\[3pt]
\hline
\hline
\cellcolor{sigstrcolor1} & $\mu_\text{ggF}^{\gamma \gamma}$ & $0.80\pm 0.19$  &1&-0.29&-0.22&0 &\cellcolor{green!50}& \multirow{4}{*}{\cite{ATLAS-CONF-2017-045,ATLAS-CONF-2017-047}} \\[3pt]
\cellcolor{sigstrcolor1} & $\mu_\text{VBF}^{\gamma \gamma}$ & $2.1\pm 0.6$  &-0.29&1&0&0&\cellcolor{green!50}&\\[3pt]
\cellcolor{sigstrcolor1} & $\mu_\text{Vh}^{\gamma \gamma}$ & $0.7\pm 0.9$  &-0.22&0&1&-0.14&\cellcolor{green!50}&\\[3pt]
\cellcolor{sigstrcolor1} & $\mu_\text{tth}^{\gamma \gamma}$ & $0.5\pm 0.6$  &0&0&-0.14&1& \cellcolor{green!50} \multirow{-4}{*}{\parbox{23pt}{36.1\\ \phantom{d}}}&\\[3pt]
\hline
\hline
\cellcolor{sigstrcolor5} & $\mu_\text{ggF}^{ZZ}$ & $1.11\pm 0.24$  &1&-0.29&-0.22&0 & \cellcolor{green!50} & \multirow{4}{*}{\cite{Aaboud:2017vzb,ATLAS-CONF-2017-047}} \\[3pt]
\cellcolor{sigstrcolor5} & $\mu_\text{VBF}^{ZZ}$ & $4.0\pm 1.6$  &-0.29&1&0&0&\cellcolor{green!50} &\\[3pt]
\cellcolor{sigstrcolor5} & $\mu_\text{Vh}^{ZZ}$ & $0\pm 1.9$  &-0.22&0&1&-0.14&\cellcolor{green!50} &\\[3pt]
\cellcolor{sigstrcolor5} & $\mu_\text{tth}^{ZZ}$ & $0\pm 3.9$  &0&0&-0.14&1&\cellcolor{green!50} \multirow{-4}{*}{\parbox{23pt}{36.1\\ \phantom{d}}}&\\[3pt]
\hline
\hline
\cellcolor{sigstrcolor4} & $\mu_\text{VBF}^{WW}$ & $1.7\pm 1.0$  & & & & & \cellcolor{red!50} 5.8 & \cite{ATLAS-CONF-2016-112} \\[3pt]
\hline
\cellcolor{sigstrcolor4} & $\mu_\text{Wh}^{WW}$ & $3.2\pm 4.3$  & & & & & \cellcolor{red!50} 5.8 & \cite{ATLAS-CONF-2016-112} \\[3pt]
\hline
\cellcolor{sigstrcolor4} & $\mu_\text{tth}^{WW}$ & $1.5\pm 0.6$  &1&-0.31&&&\cellcolor{green!50}&\multirow{2}{*}{\cite{Aaboud:2017jvq}}\\[3pt]
\cellcolor{sigstrcolor3} & $\mu_\text{tth}^{\tau \tau}$ & $1.5\pm 1.1$  &-0.31&1&&&\cellcolor{green!50} \multirow{-2}{*}{36.1}&\\[3pt]
\hline
\hline
\cellcolor{sigstrcolor2} & $\mu_\text{Vh}^{bb}$ & $1.20\pm 0.39$  & & & & & \cellcolor{green!50} 36.1 & \cite{Aaboud:2017xsd} \\[3pt]
\hline
\cellcolor{sigstrcolor2} & $\mu_\text{tth}^{bb}$ & $0.84\pm 0.63$  & & & & & \cellcolor{green!50} 36.1 & \cite{Aaboud:2017rss} \\[3pt]
\hline
\hline
\cellcolor{sigstrcolor6} & $\mu_\text{pp}^{\mu\mu}$ & $-0.1\pm 1.5$  & & & & & \cellcolor{green!50} 36.1 & \cite{Aaboud:2017ojs} \\[3pt]
\hline
\hline
& $\mu_\text{pp}^{Z\gamma}$ & $0\pm 3.4$  & & & & & \cellcolor{green!50} 36.1 & \cite{Aaboud:2017uhw} \\[3pt]
\hline
\end{tabular}
\caption{Run 2 $h$ signal strengths measured by ATLAS. Again, correlations below 0.1 were treated to be 0. The colours in the first column correspond to the ones in Figures \ref{fig:SigStr} and \ref{fig:rgagavsrg}. In the fifth column, we highlight the underlying integrated luminosity with red, yellow and green, depending on whether the measurement is based on few, moderate or the full Run 2 data.}
\label{tab:signalstrengthsA13}
\end{center}
\end{table}

\begin{table}
\begin{center}
\begin{tabular}{| l | c | c | cc | c | c |}
\hline
& \textbf{Signal} & \textbf{Value} & \multicolumn{2}{c|}{\textbf{Correlation matrix}} & ${\cal L}$ & \textbf{Source}\\[3pt]
& \textbf{strength} &  &&&\textbf{[fb$^{-1}$]}&\\[3pt]
\hline
\hline
\cellcolor{sigstrcolor1} & $\mu_\text{ggF}^{\gamma \gamma}$ & $1.11\pm 0.19$  &1&-0.32 & \cellcolor{green!50} & \multirow{2}{*}{\cite{CMS-PAS-HIG-16-040}}\\[3pt]
\cellcolor{sigstrcolor1} & $\mu_\text{VBF}^{\gamma \gamma}$ & $0.5\pm 0.6$  &-0.32&1 & \cellcolor{green!50} \multirow{-2}{*}{\parbox{23pt}{35.9\\[-7pt] \phantom{d}}}&\\[3pt]
\hline
\cellcolor{sigstrcolor1} & $\mu_\text{Vh}^{\gamma \gamma}$ & $2.3\pm 1.1$  &&& \cellcolor{green!50} 35.9 & \cite{CMS-PAS-HIG-16-040} \\[3pt]
\hline
\cellcolor{sigstrcolor1} & $\mu_\text{tth}^{\gamma \gamma}$ & $2.2\pm 0.9$  &&& \cellcolor{green!50} 35.9 & \cite{CMS-PAS-HIG-16-040} \\[3pt]
\hline
\hline
\cellcolor{sigstrcolor5} & $\mu_\text{ggF}^{ZZ}$ & $1.20\pm 0.22$  &1&-0.43& \cellcolor{green!50} & \multirow{2}{*}{\cite{Sirunyan:2017exp}}\\[3pt]
\cellcolor{sigstrcolor5} & $\mu_\text{VBF}^{ZZ}$ & $0.06\pm 1.03$  &-0.43&1& \cellcolor{green!50} \multirow{-2}{*}{\parbox{23pt}{35.9\\[-7pt] \phantom{d}}}&\\[3pt]
\hline
\cellcolor{sigstrcolor5} & $\mu_\text{Vh,h}^{ZZ}$ & $0\pm 2.85$  &&&\cellcolor{green!50} 35.9 & \cite{Sirunyan:2017exp}\\[3pt]
\hline
\cellcolor{sigstrcolor5} & $\mu_\text{Vh,l}^{ZZ}$ & $0\pm 2.78$  &&&\cellcolor{green!50} 35.9 & \cite{Sirunyan:2017exp}\\[3pt]
\hline
\cellcolor{sigstrcolor5} & $\mu_\text{tth}^{ZZ}$ & $0\pm 1.19$  &&&\cellcolor{green!50} 35.9 & \cite{Sirunyan:2017exp}\\[3pt]
\hline
\hline
\cellcolor{sigstrcolor4} & $\mu_\text{ggF}^{WW}$ & $1.02\pm 0.27$  &1&-0.24& \cellcolor{yellow!70} & \multirow{2}{*}{\cite{CMS-PAS-HIG-16-021}}\\[3pt]
\cellcolor{sigstrcolor4} & $\mu_\text{VBF+Vh}^{WW}$ & $0.89\pm 0.67$  &-0.24&1&\cellcolor{yellow!70} \multirow{-2}{*}{\parbox{23pt}{15.2\\[-7pt] \phantom{d}}}&\\[3pt]
\hline
\hline
\cellcolor{sigstrcolor3} & $\mu_\text{pp}^{\tau\tau}$ & $1.17\pm 0.44$  &&& \cellcolor{green!50} 35.9 & \cite{Sirunyan:2017khh}\\[3pt]
\hline
\cellcolor{sigstrcolor3} & $\mu_\text{ggF}^{\tau\tau}$ & $0.84\pm 0.89$  &&& \cellcolor{green!50} 35.9 & \cite{Sirunyan:2017khh}\\[3pt]
\hline
\cellcolor{sigstrcolor3} & $\mu_\text{VBF}^{\tau\tau}$ & $1.11\pm 0.35$  &&& \cellcolor{green!50} 35.9 & \cite{Sirunyan:2017khh}\\[3pt]
\hline
\cellcolor{sigstrcolor3} & $\mu_\text{tth}^{\tau\tau}$ & $0.72\pm 0.58$  &&& \cellcolor{green!50} 35.9 & \cite{CMS-PAS-HIG-17-003}\\[3pt]
\hline
\hline
\cellcolor{sigstrcolor2} & $\mu_\text{VBF}^{bb}$ & $-3.7\pm 2.5$  &&& \cellcolor{red!50} 2.3 & \cite{CMS-PAS-HIG-16-003}\\[3pt]
\hline
\cellcolor{sigstrcolor2} & $\mu_\text{Vh}^{bb}$ & $1.2\pm 0.4$  &&& \cellcolor{green!50} 35.9 & \cite{Sirunyan:2017elk}\\[3pt]
\hline
\cellcolor{sigstrcolor2} & $\mu_\text{tth}^{bb}$ & $-0.19\pm 0.81$  &&& \cellcolor{yellow!70} 12.9 & \cite{CMS-PAS-HIG-16-038}\\[3pt]
\hline
\hline
\cellcolor{sigstrcolor6} & $\mu_\text{pp}^{\mu\mu}$ & $0.7\pm 1.0$  &&& \cellcolor{green!50} 35.9 & \cite{CMS-PAS-HIG-17-019} \\[3pt]
\hline
\end{tabular}
\caption{Run 2 $h$ signal strengths measured by CMS. The colours in the first column correspond to the ones in Figures \ref{fig:SigStr} and \ref{fig:rgagavsrg}, the ones in the fifth column to the amount of underlying data like in Table \ref{tab:signalstrengthsA13}.}
\label{tab:signalstrengthsC13}
\end{center}
\end{table}

A list of the available experimental signal strength values from LHC Run 1 and 2 can be found in the Tables \ref{tab:signalstrengths8} (ATLAS and CMS combination for Run 1), \ref{tab:signalstrengthsA13} (ATLAS numbers for Run 2) and \ref{tab:signalstrengthsC13} (CMS measurements for Run 2). For the Run 2 data, we also list the corresponding integrated luminosities ${\cal L}$. The numbers for the correlations in Table \ref{tab:signalstrengths8} can be found in the mentioned document. For Run 2, ATLAS provides correlations only for the combination of the $\gamma \gamma$ and $ZZ$ decays; observing very similar numbers in the corresponding Run 1 data, we assume identical correlations for the $\gamma \gamma$ and $ZZ$ final states. The correlation between $\mu_\text{tth}^{WW}$ and $\mu_\text{tth}^{\tau \tau}$ was extracted from Figure 17 in \cite{Aaboud:2017jvq}. (We assume that the $VV$ final state therein is dominated by $WW$.) Also the CMS correlations in Table \ref{tab:signalstrengthsC13} were reconstructed from the signal strength contours (or cross section times branching ratio contours) in the plane of VBF vs.~ggF production.
In Section \ref{sec:signalstrengths} we discuss the individual impact of the signal strengths on the 2HDM parameters, ordered by the decay products.

\begin{table}
\begin{center}
\begin{small}
\begin{tabular}{| l | l | l | l c | c | c |}
\hline
\multicolumn{2}{| l |}{\textbf{Label}} &\textbf{Channel} & \textbf{Experiment} && \textbf{Mass range} & ${\cal L}$ \\
\multicolumn{2}{| l |}{}&&&& \textbf{[GeV]} & \textbf{[fb$^{-1}$]} \\[1pt]
\hline
\hline
$C_{8b}^{bb}$ & \cellcolor{color2} &$bb \to H/A \to bb$ & CMS & \cite{Khachatryan:2015tra} & [100;900] & 19.7 \\
\hline
\hline
$A_{8}^{\tau\tau}$ & \cellcolor{color5} &\multirow{2}{*}{$gg\to H/A \to \tau\tau$} & ATLAS &\cite{Aad:2014vgg} & [90;1000] & 19.5-20.3 \\
$C_{8}^{\tau\tau}$ & \cellcolor{color6} & & CMS &\cite{CMS:2015mca} &  [90;1000]  &19.7 \\
\hline
$A_{8b}^{\tau\tau}$ & \cellcolor{color7} &\multirow{2}{*}{$bb\to H/A \to \tau\tau$} & ATLAS &\cite{Aad:2014vgg} & [90;1000] & 19.5-20.3 \\
$C_{8b}^{\tau\tau}$ & \cellcolor{color8} & & CMS & \cite{CMS:2015mca}& [90;1000] & 19.7 \\
\hline
\hline
$A_{8}^{\gamma\gamma}$ & \cellcolor{color3} &$gg\to H/A \to \gamma\gamma$ & ATLAS &\cite{Aad:2014ioa} & [65;600] & 20.3 \\
\hline
\hline
$A_{8}^{Z\gamma}$ & \cellcolor{color3} &\multirow{2}{*}{$pp\to H/A \to Z\gamma \to (\ell \ell) \gamma$} & ATLAS & \cite{Aad:2014fha} & [200;1600] & 20.3 \\
$C_{8}^{Z\gamma}$ & \cellcolor{color4} & & CMS & \cite{CMS:2016all} & [200;1200] & 19.7 \\
\hline
\hline
$A_{8}^{ZZ}$ & \cellcolor{color10} &$gg\to H\to ZZ$ & ATLAS & \cite{Aad:2015kna}& [140;1000] & 20.3 \\
$A_{8V}^{ZZ}$ & \cellcolor{color11} &$VV \to H\to ZZ$ & ATLAS & \cite{Aad:2015kna}& [140;1000] & 20.3 \\
\hline
\hline
$A_{8}^{WW}$ & \cellcolor{color7} &$gg\to H\to WW$ & ATLAS &\cite{Aad:2015agg}& [300;1500] & 20.3 \\
$A_{8V}^{WW}$ & \cellcolor{color8} &$VV \to H\to WW$ & ATLAS & \cite{Aad:2015agg}& [300;1500] & 20.3 \\
\hline
\hline
$C_{8}^{VV}$ & \cellcolor{color9} & $pp \to H\to VV$ & CMS & \cite{Khachatryan:2015cwa} & [145;1000] & 24.8 \\
\hline
\hline
$C_{8}^{4b}$ & \cellcolor{color9} &$pp\to H\to hh \to (bb) (bb)$ & CMS &\cite{Khachatryan:2015yea} & [270;1100] & 17.9\\
$C_{8}^{2\gamma2b}$ & \cellcolor{color10} &$pp\to H\to hh \to (\gamma \gamma) (bb)$ & CMS & \cite{Khachatryan:2016sey} & [260;1100] & 19.7\\
$A_{8}^{hh}$ & \cellcolor{color11} &$gg\to H\to hh$ & ATLAS &\cite{Aad:2015xja} & [260;1000] & 20.3\\
$C_{8}^{2b2\tau}$ & \cellcolor{color12} &$pp\to H\to hh [\to (bb) (\tau\tau)]$ & CMS & \cite{Sirunyan:2017tqo} & [300;1000] & 18.3\\
\hline
\hline
$A_{8}^{bbZ}$ & \cellcolor{color3} &$gg\to A\to hZ \to (bb) Z$ & ATLAS & \cite{Aad:2015wra} & [220;1000] & 20.3\\
$A_{8}^{\tau\tau Z}$ & \cellcolor{color4} &$gg\to A\to hZ \to (\tau\tau) Z$ & ATLAS & \cite{Aad:2015wra} & [220;1000] & 20.3\\
$C_{8}^{2b2\ell}$ & \cellcolor{color5} &$gg\to A\to hZ \to (bb) (\ell \ell)$ & CMS &\cite{Khachatryan:2015lba} & [225;600] &19.7\\
$C_{8}^{2\tau2\ell}$ & \cellcolor{color6} &$gg\to A\to hZ \to (\tau\tau) (\ell \ell)$ & CMS & \cite{Khachatryan:2015tha} & [220;350] & 19.7\\
\hline
\hline
$C_{8}^{AZ}$ & & $pp\to H\to AZ \to (bb) (\ell\ell)$ & CMS
 & \cite{Khachatryan:2016are} & [130;1000] & 19.8\\
\hline
$C_{8}^{HZ}$ & & $pp\to A\to HZ \to (bb) (\ell\ell)$ & CMS
 & \cite{Khachatryan:2016are} & [130;1000] & 19.8\\
\hline
\hline
$A_{8}^{\tau\nu}$ & \cellcolor{color3} &$pp\to H^\pm \to \tau^\pm \nu $ & ATLAS &\cite{Aad:2014kga} & [180;1000] & 19.5\\
$C_{8}^{\tau\nu}$ & \cellcolor{color4} &$pp\to H^+ \to \tau^+ \nu $ & CMS &\cite{Khachatryan:2015qxa}& [180;600] & 19.7\\
\hline
\hline
$A_{8}^{tb}$ & \cellcolor{color2} &$pp\to H^\pm \to t b $ & ATLAS & \cite{Aad:2015typ} & [200;600] & 20.3\\
$C_{8}^{tb}$ & \cellcolor{color3} &$pp\to H^+ \to t \bar{b} $ & CMS & \cite{Khachatryan:2015qxa} & [180;600] & 19.7\\
\hline
\end{tabular}
\caption{List of the available heavy Higgs searches from LHC Run 1 relevant for the 2HDM. In the first column, we assign a label and colour to each search, which correspond to the ones in Figures \ref{fig:Htobb} to \ref{fig:Hptotaunu}. Details of production and decay modes are given in the second column. The third column contains the corresponding reference. The mass ranges, for which the corresponding limits on \sigbr are given, and the integrated luminosity the searches are based on, can be found in the fourth and fifth column. The CMS Run 1 limits of the di-photon channel are included in their Run 2 bounds. $VV$ refers to either $WW$ or $ZZ$. $C_{8}^{VV}$ provides signal strength limits. $A_{8}^{hh}$ contains information about the decays of $hh$ to $4b$, $2\tau 2b$, $2\gamma 2b$ and $2\gamma 2W$.}
\label{tab:8TeV}
\end{small}
\end{center}
\end{table}

\begin{table}
\begin{center}
\begin{small}
\begin{tabular}{| l | l | l | l c | c | c |}
\hline
\multicolumn{2}{| l |}{\textbf{Label}} &\textbf{Channel} & \textbf{Experiment} && \textbf{Mass range} & ${\cal L}$ \\
\multicolumn{2}{| l |}{}&&&& \textbf{[GeV]} & \textbf{[fb$^{-1}$]} \\[1pt]
\hline
\hline
$C_{13}^{bb}$ & \cellcolor{color1} &$pp \to H/A\to bb$ & CMS
 & \cite{CMS-PAS-HIG-16-025} & [0.55;1.2] & \cellcolor{red!50} 2.69\\
\hline
\hline
$A_{13}^{\tau\tau}$ & \cellcolor{color1} &\multirow{2}{*}{$gg \to H/A\to \tau \tau$} & ATLAS
 & \cite{Aaboud:2017sjh} & [0.2;2.25] & \cellcolor{green!50} 36.1\\[-1pt]
$C_{13}^{\tau\tau}$ & \cellcolor{color2} & & CMS
 & \cite{CMS-PAS-HIG-16-037} & [0.09;3.2] & \cellcolor{yellow!70} 12.9\\
\hline
$A_{13b}^{\tau\tau}$ & \cellcolor{color3} &\multirow{2}{*}{$bb \to H/A\to \tau \tau$} & ATLAS
 & \cite{Aaboud:2017sjh} & [0.2;2.25] & \cellcolor{green!50} 36.1\\[-1pt]
$C_{13b}^{\tau\tau}$ & \cellcolor{color4} & & CMS
 & \cite{CMS-PAS-HIG-16-037} & [0.09;3.2] & \cellcolor{yellow!70} 12.9\\
\hline
\hline
$A_{13}^{\gamma\gamma}$ & \cellcolor{color1} &$pp \to H/A\to \gamma \gamma$ & ATLAS & \cite{Aaboud:2017yyg} & [0.2;2.7] & \cellcolor{green!50} 36.7\\
\hline
$C_{13}^{\gamma\gamma}$ & \cellcolor{color2} & $gg \to H/A\to \gamma \gamma$ & CMS
 & \cite{CMS-PAS-EXO-16-027} & [0.5;4] & \cellcolor{green!50} 35.9\\ 
\hline
\hline
$A_{13}^{Z\gamma}$ & \cellcolor{color1} & $gg \to H/A\to Z \gamma [\to (\ell \ell) \gamma ]$ & ATLAS
 & \cite{Aaboud:2017uhw} & [0.25;2.4] & \cellcolor{green!50} 36.1\\
\hline
$C_{13}^{Z\gamma}$ & \cellcolor{color2} & $gg \to H/A\to Z \gamma$ & CMS
 & \cite{Sirunyan:2017hsb} & [0.35;4] & \cellcolor{green!50} 35.9\\
\hline
\hline
$A_{13}^{2\ell2L}$ & \cellcolor{color1} & $gg\to H \to ZZ [\to (\ell \ell) (\ell \ell, \nu \nu)]$ & ATLAS
 & \cite{Aaboud:2017rel} & [0.2;1.2] & \cellcolor{green!50} 36.1\\
\hline
$A_{13V}^{2\ell2L}$ & \cellcolor{color2} & $VV\to H \to ZZ [\to (\ell \ell) (\ell \ell, \nu \nu)]$ & ATLAS
 & \cite{Aaboud:2017rel} & [0.2;1.2] & \cellcolor{green!50} 36.1\\
\hline
$C_{13}^{2\ell2\nu}$ & \cellcolor{color3} & $pp\to H \to ZZ [\to (\ell \ell) (\nu \nu)]$ & CMS
 & \cite{CMS-PAS-B2G-16-023} & [0.6;2.5] & \cellcolor{green!50} 35.9\\
\hline
$C_{13g}^{2\ell2\nu}$ & \cellcolor{color4} & $gg\to H \to ZZ [\to (\ell \ell) (\nu \nu)]$ & CMS
 & \cite{CMS-PAS-HIG-16-001} & [0.2;0.6] & \cellcolor{red!50} 2.3\\
\hline
$C_{13V}^{2\ell2\nu}$ & \cellcolor{color5} & $VV\to H \to ZZ [\to (\ell \ell) (\nu \nu)]$ & CMS
 & \cite{CMS-PAS-HIG-16-001} & [0.2;0.6] & \cellcolor{red!50} 2.3\\
\hline
$C_{13V}^{4\ell}$ & \cellcolor{color6} & $(VV+VH)\to H\to ZZ\to (\ell \ell)(\ell \ell)$ & CMS
 & \cite{CMS-PAS-HIG-16-033} & [0.13;2.53] & \cellcolor{yellow!70} 12.9\\
\hline
$C_{13}^{2\ell 2q}$ & \cellcolor{color7} & $pp\to H\to ZZ [\to (\ell \ell) (qq)]$ & CMS
 & \cite{CMS-PAS-HIG-16-034} & [0.5;2] & \cellcolor{yellow!70} 12.9\\
\hline
$A_{13}^{2L2q}$ & \cellcolor{color8} & $gg\to H\to ZZ [\to (\ell \ell, \nu \nu) (qq)]$ & ATLAS
 & \cite{Aaboud:2017itg} & [0.3;3] & \cellcolor{green!50} 36.1\\
\hline
$A_{13V}^{2L2q}$ & \cellcolor{color9} & $VV\to H\to ZZ [\to (\ell \ell, \nu \nu) (qq)]$ & ATLAS
 & \cite{Aaboud:2017itg} & [0.3;3] & \cellcolor{green!50} 36.1\\
\hline
\hline
$A_{13}^{2(\ell\nu)}$ & \cellcolor{color1} & $gg\to H\to WW [\to (e \nu) (\mu \nu)]$ & ATLAS
 & \cite{Aaboud:2017gsl} & [0.25;4] & \cellcolor{green!50} 36.1\\
\hline
$A_{13V}^{2(\ell\nu)}$ & \cellcolor{color2} & $VV\to H\to WW [\to (e \nu) (\mu \nu)]$ & ATLAS
 & \cite{Aaboud:2017gsl} & [0.25;3] & \cellcolor{green!50} 36.1\\
\hline
$C_{13}^{2(\ell\nu)}$ & \cellcolor{color3} & $(gg\!+\!VV)\to H\to WW \to (\ell \nu) (\ell \nu)$ & CMS
 & \cite{CMS-PAS-HIG-16-023} & [0.2;1] & \cellcolor{red!50} 2.3\\
\hline
$A_{13}^{\ell\nu2q}$ & \cellcolor{color4} & $gg\to H\to WW[\to (\ell \nu) (qq)]$ & ATLAS
 & \cite{Aaboud:2017fgj} & [0.3;3] & \cellcolor{green!50} 36.1\\
\hline
$A_{13V}^{\ell\nu2q}$ & \cellcolor{color5} & $VV\to H\to WW[\to (\ell \nu) (qq)]$ & ATLAS
 & \cite{Aaboud:2017fgj} & [0.3;3] & \cellcolor{green!50} 36.1\\
\hline
\hline
$A_{13}^{4q}$ & \cellcolor{color6} & $pp\to H\to VV [\to (qq) (qq)]$ & ATLAS
 & \cite{Aaboud:2017eta} & [1.2;3] & \cellcolor{green!50} 36.7\\
\hline
\hline
$A_{13}^{4b}$ & \cellcolor{color1} & \multirow{2}{*}{$pp \to H\to hh \to (bb) (bb)$} & ATLAS
 & \cite{ATLAS-CONF-2016-049} & [0.3;3]  & \cellcolor{yellow!70} 13.3\\[-1pt]
$C_{13}^{4b}$ & \cellcolor{color2} & & CMS
 & \cite{CMS-PAS-HIG-17-009} & [0.26;1.2] & \cellcolor{green!50} 35.9\\
\hline
$C_{13g}^{4b}$ & \cellcolor{color3} & $gg \to H\to hh \to (bb) (bb)$ & CMS
 & \cite{Sirunyan:2017isc} & [1.2;3]  & \cellcolor{green!50} 35.9\\ 
\hline
$A_{13}^{2\gamma2b}$ & \cellcolor{color4} & $pp \to H\to hh [\to (\gamma \gamma) (bb)]$ & ATLAS
 & \cite{ATLAS-CONF-2016-004} & [0.275;0.4] & \cellcolor{red!50} 3.2\\
$C_{13}^{2\gamma2b}$ & \cellcolor{color5} & $pp \to H\to hh \to (\gamma \gamma) (bb)$ & CMS
 & \cite{CMS-PAS-HIG-17-008} & [0.25;0.9] & \cellcolor{green!50} 35.9\\
\hline
$C_{13}^{2b2\tau}$ & \cellcolor{color6} & $pp \to H\to hh \to (bb) (\tau \tau)$ & CMS
 & \cite{Sirunyan:2017djm} & [0.25;0.9] & \cellcolor{green!50} 35.9\\
\hline
$C_{13}^{2b2V}$ & \cellcolor{color7} & $pp \to H\to hh \to (bb) (VV\to \ell \nu \ell \nu)$ & CMS
 & \cite{Sirunyan:2017guj} & [0.26;0.9] & \cellcolor{green!50} 36\\
\hline
$A_{13}^{2\gamma2W}$ & \cellcolor{color8} & $gg \to H\to hh [\to (\gamma \gamma) (WW
)]$ & ATLAS
 & \cite{ATLAS-CONF-2016-071} & [0.25;0.5] & \cellcolor{yellow!70} 13.3\\
\hline
\hline
$A_{13}^{bbZ}$ & \cellcolor{color1} & $gg\to A\to hZ \to (bb) Z$ & ATLAS
 & \cite{Aaboud:2017cxo} & [0.2;2] & \cellcolor{green!50} 36.1\\
\hline
$A_{13b}^{bbZ}$ & \cellcolor{color2} & $b\bar{b}\to A\to hZ \to (bb) Z$ & ATLAS
 & \cite{Aaboud:2017cxo} & [0.2;2] & \cellcolor{green!50} 36.1\\
\hline
\end{tabular}
\caption{List of the available neutral heavy Higgs searches from LHC Run 2 relevant for the 2HDM. For an explanation, see the description below Table \ref{tab:8TeV}. In the last column, we additionally highlight an underlying integrated luminosity of around 3, 13 or 36 fb$^{-1}$ in red, yellow or green, respectively.}
\label{tab:13TeV}
\end{small}
\end{center}
\end{table}

\begin{table}
\begin{center}
\begin{small}
\begin{tabular}{| l | l | l | l c | c | c |}
\hline
\multicolumn{2}{| l |}{\textbf{Label}} &\textbf{Channel} & \textbf{Experiment} && \textbf{Mass range [TeV]} & ${\cal L}$ \textbf{[fb$^{-1}$]}\\
\hline
\hline
$A_{13}^{\tau\nu}$ & \cellcolor{color1} & \multirow{2}{*}{$pp\to H^{\pm} \to \tau^\pm \nu $} & ATLAS
 & \cite{ATLAS-CONF-2016-088} & [0.2;2] & \cellcolor{yellow!70} 14.7\\
$C_{13}^{\tau\nu}$ & \cellcolor{color2} & & CMS
 & \cite{CMS-PAS-HIG-16-031} & [0.18;3] & \cellcolor{yellow!70} 12.9\\
\hline
\hline
$A_{13}^{tb}$ & \cellcolor{color1} & \multirow{2}{*}{$pp\to H^+ \to t\bar b$} & ATLAS
 & \cite{ATLAS-CONF-2016-089} & [0.3;1] & \cellcolor{yellow!70} 13.2\\
$A_{13}^{tb}$ & \cellcolor{color1} & & ATLAS
 & \cite{ATLAS-CONF-2016-104} & [0.2;0.3]$\cup$[1;2] & \cellcolor{yellow!70} 13.2\\
\hline
\end{tabular}
\caption{List of the available charged heavy Higgs searches from LHC Run 2 relevant for the 2HDM. For an explanation, see the description below Table \ref{tab:8TeV}.}
\label{tab:13TeVcharged}
\end{small}
\end{center}
\end{table}

Concerning the direct searches for the heavy CP-even, the CP-odd and the charged Higgs, we have updated the number of used LHC analyses from 16 in \cite{Cacchio:2016qyh} to 50 Run 1 and 2 measurements in the present article. We calculate the product of the production cross section \cite{LHCHXSWG,Spira:1995mt,Harlander:2012pb,Alwall:2014hca,Degrande:2016hyf,Degrande:2015vpa,Flechl:2014wfa,deFlorian:2016spz,Dittmaier:2009np,Berger:2003sm} and the branching ratio \cite{Djouadi:1997yw,Agashe:2014kda} of a specific decay, \sigbr. In order to compare it with the experimental bounds, we assign Gaussian likelihoods with a central value of 0 to the ratio of the theoretical value and the observed upper limit of \sigbr. This method agrees with the treatment of the upper limit of the $Z\gamma$ signal strength mentioned above and coincides with our approach in \cite{Cacchio:2016qyh} under the assumption that the observed upper limit does not deviate from the expected one. With no evidence for such a deviation in any of the searches, this approximation seems to be justified. The experimental input from Run 1 and 2 can be found in Tables \ref{tab:8TeV}, \ref{tab:13TeV} and \ref{tab:13TeVcharged}. These analyses comprise a large variety of searches for heavy resonances decaying into fermionic or bosonic states: $bb$, $\tau\tau$, $\gamma\gamma$ and $Z\gamma$ limits can be applied to both, the CP-even and CP-odd Higgs bosons; signatures with a pair of massive bosons or AZ in the final state can exclusively stem from $H$ decays at tree-level in the 2HDM, while a CP-odd resonance decaying to one $h$ or $H$ and one $Z$ is interpreted as $A$; finally, the searches for charged Higgs particles were performed looking for the final states $tb$ or $\tau\nu$. If the branching ratio into a specific final state -- like for instance $(\gamma\gamma) (bb)$ -- is included in the upper limit, we list it in the table. If the final state is not included in the \sigbr limits, but its information is needed to distinguish it from other searches, we write it in square brackets. The secondary decay products of one particle are combined in parentheses. In the case in which two concurring searches are available that are partially based on the same set of data, we use the limit which is derived from the larger amount of data. For instance, the latest CMS results of the searches for an $H$ decaying via $ZZ$ into two leptons and two neutrinos are available only for $m_H>600$ GeV. Lighter $m_H$ scenarios will be constrained using an older publication based on an integrated luminosity of $2.3$ fb$^{-1}$. Also for the upper limits on $H\to hh \to (b\bar{b}) (b\bar{b})$ by CMS and on $H^+\to t\bar{b}$ by ATLAS we apply different searches depending on the masses.
For $gg \to X\to \gamma \gamma$, CMS combined their 8 and 13 TeV data; the limits are given for the 13 TeV production cross section.
A detailed discussion of how the different searches constrain the 2HDM can be found in Section \ref{sec:heavy}, where we show the results ordered by the decay products.

Apart from the discussed tree-level Higgs observables the 2HDM scalars can also contribute to the quantum corrections of other observables, the most important ones being the electroweak precision observables, the $b\to s \gamma$ branching ratio and the mass difference in the $B_s$ meson system. While the implementation into \HEPfit was already explained in \cite{Cacchio:2016qyh}, we updated the experimental values \cite{deBlas:2017wmn,HFLAVhomepage,Amhis:2016xyh}.
Also for the treatment of theoretical constraints we refer to \cite{Cacchio:2016qyh}, with two exceptions: We do not apply any constraints arising from the renormalization group evolution and define our model at the electroweak scale. And for the next-to-leading order unitarity bounds we chose the most conservative approach that appeared reasonable to us, namely requiring that the real and imaginary parts of the $S$-matrix eigenvalues should be between $-0.5$ and $0.5$ and between $0$ and $1$, respectively. Moreover we impose perturbativity by discarding scenarios for which the one-loop contribution to these eigenvalues exceeds the tree-level term in magnitude.

As numerical set-up we use the open-source package \HEPfit \cite{hepfit}, interfaced with the release candidate of the Bayesian Analysis Toolkit (BAT) \cite{Caldwell:2008fw}. The former calculates all mentioned 2HDM observables and feeds them into the parallelized BAT, which applies the Bayesian fit with Markov chain Monte Carlo simulations.

\section{$h$ signal strengths}
\label{sec:signalstrengths}

In this section we show the impact of the $h$ signal strengths on the 2HDM parameters. The fits were done with the most up-to-date experimental inputs; for a comparison with the status before EPS-HEP 2017, see \cite{Eberhardt:2017ulj}.
The differences in the $\mathbb{Z}_2$ symmetry assignment to the fermions result in a type dependent treatment of their couplings to the light Higgs boson.
The signal strength of the process with a given initial state $i$ producing an $h$ which decays to the final state $f$ can be written as
\begin{align}
 \mu _i^f &= r_i \cdot \frac{r_f}{\sum \limits_{f^\prime} r_{f^\prime} {\cal B}_\text{\tiny SM}(h\to f^\prime)}, \label{eq:signalstrengths}
\end{align}
where $r_x$ is the ratio of the 2HDM and the SM partial width of an $h$ decaying into $x$ and ${\cal B}_\text{\tiny SM}(h\to x)$ is the corresponding SM branching ratio.
From this equation one can see that every signal strength depends on the 2HDM $h$ couplings of all decay products.

\begin{figure}
   \begin{picture}(450,350)(0,0)
    \put(30,0){\includegraphics[width=350pt]{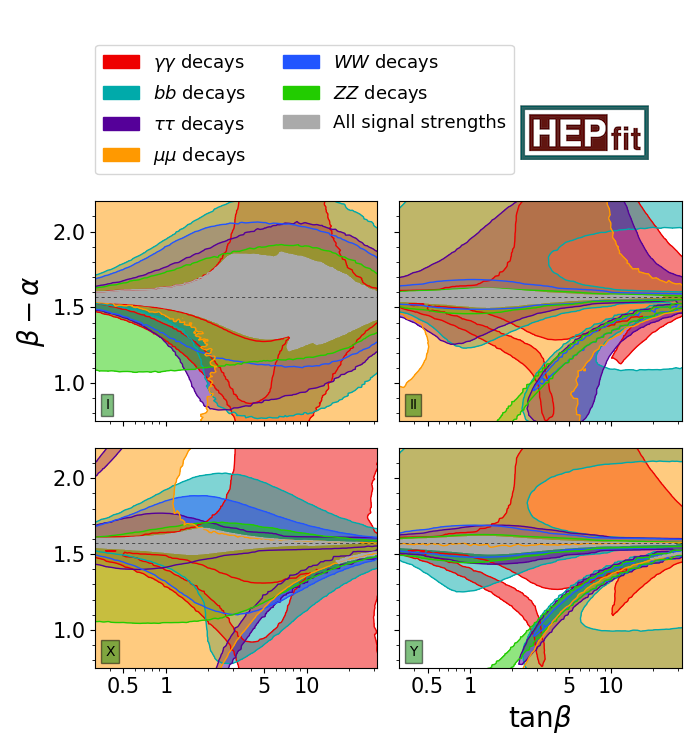}}
   \end{picture}
  \caption{The impact of the $h$ signal strength measurements is illustrated in the $\beta-\alpha$ vs.~$\tan \beta$ plane in all four 2HDM types. We show the 95.4\% posterior probability contours for individual fits to data from $h$ decays to $\gamma\gamma$, $bb$, $\tau\tau$, $\mu\mu$, $WW$ and $ZZ$ in red, cyan, purple, orange, blue and green, respectively. The resulting 95.4\% regions of the combined fits to all signal strengths are the grey areas.}
  \label{fig:SigStr}
\end{figure}

In Figure \ref{fig:SigStr} we show the individual impact of the signal strengths with a specific final state on the $\beta-\alpha$ vs.~$\tan \beta$ plane as well as their combination in all four types of $\mathbb{Z}_2$ symmetry. We have tried to adopt the colouring scheme from Figure 14 of the Run 1 combination \cite{Khachatryan:2016vau}. All contours delimit the regions allowed with a probability of 95.4\%. The upper limit on $\mu ^{Z\gamma}_{pp}$ is included in the combination, but not shown separately as its effect is minimal.

In type I all fermions have the same relative coupling to $h$: $r_{tt}=r_{bb}=r_{\tau\tau}=r_{\mu\mu}=\cos(\beta-\alpha)/\tan\beta+\sin(\beta-\alpha)$. This can only deviate significantly from 1 if $\tan \beta$ is smaller than 1 and $\beta-\alpha$ is not close to the alignment limit $\pi/2$. In these regions, the di-photon signal strengths are the most constraining ones, see the upper left panel of Figure \ref{fig:SigStr}. For larger $\tan \beta$ values, the $ZZ$ and $WW$ signal strengths become the most important constraints. In the combined fit to all signal strengths, the largest possible deviation of $\beta-\alpha$ from $\pi/2$ is $0.26$ at 95.4\% if we marginalize over all other parameters.

The upper right panel of Figure \ref{fig:SigStr} shows fit results with the same inputs for type II. Here, the relative down-type fermion and lepton couplings to $h$ are different from the top coupling, and thus the fermionic signal strengths yield more powerful constraints. But also the signal strengths with a bosonic final state become stronger because of the modifications of the loop coupling $r_{gg}$ and the fermionic couplings in the denominator of Eq.~\eqref{eq:signalstrengths}. Especially the $WW$ and $ZZ$ signal strengths constrain $\beta-\alpha$ to be very close to $\pi/2$; the largest deviation from the alignment limit in the one-dimensional fit to all signal strengths is $0.055$ at 95.4\%. The so-called ``wrong-sign'' solution for the fermionic couplings, which is represented by the lower ``branches'' of the individual $WW$, $ZZ$, $bb$ and $\tau\tau$ signal strength fits for $\tan\beta>3$ and $\beta-\alpha<1.5$, can (cannot) be excluded in the combined fit to all signal strengths with a probability of $95.4\%$ ($99.7\%$). These scenarios have also been shown to be incompatible with the assumption that the 2HDM of type II is stable under a renormalization group evolution up to ${\cal O}(1)$ TeV \cite{Cacchio:2016qyh,Basler:2017nzu}.

\begin{figure}
   \begin{picture}(450,350)(0,0)
    \put(30,0){\includegraphics[width=350pt]{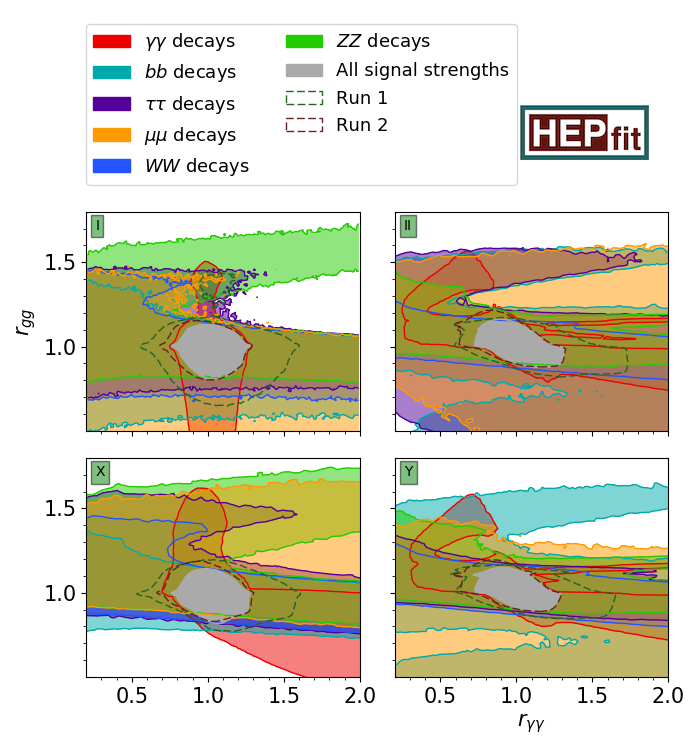}}
   \end{picture}
  \caption{The $95.4\%$ probability contours for different combinations of signal strengths are shown in the plane of the relative one-loop couplings of the $h$ to gluons and photons. While the colours of the shaded contours with solid borders correspond to the ones in Figure \ref{fig:SigStr}, the Run 1 and Run 2 combinations are bounded by the dark green and brown dashed contours, respectively.}
  \label{fig:rgagavsrg}
\end{figure}

In type X, the $h$ couplings of the down-type quarks agree with the ones of the top quark, but the leptonic couplings are like in type II. Consequently, the contour of the $bb$ decays in the lower left panel of Figure \ref{fig:SigStr} has a similar shape as the one of type I, while the $\tau\tau$ and $\mu\mu$ decays behave more like in type II for large $\tan\beta$ for $\beta-\alpha<\pi/2$. For $\tan\beta>2$ the latter two are the dominant signal strengths. For very large $\tan\beta$, the wrong-sign solution of the fermion couplings is allowed at $95.4\%$. However, no larger deviations of $\beta-\alpha$ from $\pi/2$ than $0.069$ are allowed at the $95.4\%$ level if we combine all signal strength information and marginalize over all other parameters.

Finally, the type Y fit can be found in the lower right panel of Figure \ref{fig:SigStr}. Like in type II, $\beta-\alpha$ has to be very close to the alignment limit with the bosonic signal strengths being the strongest constraints. But like in type X, the wrong-sign coupling of the fermions cannot be completely excluded at $95.4\%$ in the fit combining all signal strengths, although it is only possible for very large $\tan\beta$. In this type's combined fit and marginalizing over the other parameters, $\beta-\alpha$ cannot be further away from $\pi/2$ than $0.056$ with a probability of $95.4\%$.

As compared to the status before EPS-HEP 2017 \cite{Eberhardt:2017ulj}, the $WW$, $\gamma\gamma$ and $bb$ signal strengths have become more constraining; the latter changed drastically due to additionally released data. In type II, a small spot of the ``wrong sign'' branch around $\tan \beta=3$ and $\beta-\alpha=1$ was allowed at the $95.4\%$ before summer 2017 and has disappeared now.

The two angles $\alpha$ and $\beta$ define all tree-level couplings of fermions and bosons to the light Higgs $h$, but the loop couplings to gluons and photons are more complicated. In order to analyse their allowed ranges, we show the $r_{gg}$ vs.~$r_{\gamma\gamma}$ plane in Figure \ref{fig:rgagavsrg}. Apart from the individual fits to the different final states and their combination like in Figure \ref{fig:SigStr}, we also add the contours from a fit to only Run 1 and only Run 2 data, respectively.

In all types, the combined fit to all signal strengths is dominated by the bosonic decays. While the $\mu^{\gamma\gamma}$ mainly delimit $r_{\gamma\gamma}$, $r_{gg}$ is constrained also by the $\mu^{WW}$ and $\mu^{ZZ}$ measurements. The maximal deviation of $r_{\gamma\gamma}$ ($r_{gg}$) from its SM value is roughly $30\%$ ($20\%$).
The wrong-sign solution for the fermion couplings can be seen in type II, X and Y: The regions for $r_{gg}>1$ in Figure \ref{fig:rgagavsrg} contain the lower branches of Figure \ref{fig:SigStr}. For type II, it has been shown that the wrong-sign couplings feature increased $r_{gg}$ and reduced $r_{\gamma\gamma}$ \cite{Eberhardt:2013uba}. In the lower right panel of Figure \ref{fig:rgagavsrg}, this ``second solution'' is visible between $1.1$ and $1.2$ for $r_{gg}$ as spikes in the $WW$, $ZZ$ and $\tau\tau$ contours for large $r_{\gamma\gamma}$ as well as in the combined signal strength fit in both $r_{\gamma\gamma}$ directions.
Comparing all Run 1 signal strengths with all Run 2 signal strengths, one can see that generally the Run 2 data is more constraining in all types. However, the Run 1 signal strengths prefer a smaller $r_{gg}$ and thus determine the upper limit of the gluon coupling ratio in the combined fit to all signal strengths. This can be seen in the types II and Y.

The run time for fits with 120 million iterations was $230\pm 80$ CPU hours for the di-photon final state and $70\pm 20$ CPU hours for the other single channels.

\section{Heavy Higgs searches}
\label{sec:heavy}

In the following, we will scrutinize the impact of the searches for heavy Higgs particles in all four types of a 2HDM with a softly broken $\mathbb{Z}_2$ symmetry, ordered by their decay products. First we will address the fermionic decays to $tt$, $bb$ and $\tau\tau$ and the loop induced decays with $\gamma \gamma$ and $Z\gamma$ in the final state. The searches for signals in these channels apply to both, $H$ and $A$ bosons. After that, we will turn towards the $H$ specific decays into two massive vector bosons or two $h$ bosons and the $A$ specific channel with an $h$ and a $Z$ in the final state. Finally, the decays of a charged Higgs to $\tau\nu$ and $tb$ will be discussed.
The limits $C_{8}^{AZ}$ and $C_{8}^{HZ}$ on the decays $H\to AZ$ and $A\to HZ$ were only used in the combination of all heavy searches; their impact on the $m_A$ vs.~$m_H$ plane is found to be weak if we marginalize over all other parameters.
The narrow width approximation will be applied throughout this section; we will comment on its validity at the end of the next section.

The grey shaded regions in the plots in this section depict the prediction of the 13 TeV \sigbr for the corresponding channel \textit{without} applying any theoretical or experimental constraints on the model; in other words they correspond to our priors. The black dashed lines delimit the available ranges of the \sigbr for the corresponding channel when \textit{only} the theoretical constraints defined in the Sec.~\ref{sec:constraints} have been used in the fit. The areas within the various coloured solid lines depict the $95.4$\% posterior ranges of \sigbr after imposing the experimental constraints from the LHC for a particular measurement. The legend of each plot refers to the channels described in Tables \ref{tab:8TeV}, \ref{tab:13TeV} and \ref{tab:13TeVcharged}. The horizontal coloured lines on the top of the panels mark the mass ranges analyzed at the LHC for each of the searches denoted in the legend. In the following plots the posterior prediction of \sigbr after considering a particular direct search replicates the prior behaviour unless it deviates from rest of the posteriors for the same channel. In other words, the direct searches only have a visible impact on the 2HDM parameters if their contour is lower than the other coloured contours. Also the lower lines of the searches mostly represent the priors and do not pose any significant lower limit on \sigbr.

The fits to the single experiments in this section took $60\pm 7$ CPU hours with 120 million iterations.

\subsection{$H$ and $A$ decays to $t\bar{t}$}
\label{sec:resultst}

For $H$ and $A$ masses heavier than two top quarks the decay to $t\bar{t}$ is the dominant in the 2HDM, at least for moderate values of $\tan \beta$. Unfortunately, a possible signal strongly interferes with the tree-level background process $gg\to t\bar{t}$.
The only available experimental limits of an analysis which takes into account this interference is \cite{Aaboud:2017hnm} for the 2HDM of type II. Its limits, however, constrain only a small region for $\tan \beta \approx 1$ and $m_{H/A}\approx 500$ GeV, which has been excluded by indirect constraints. Therefore, we do not take into account this direct measurement.

\subsection{$H$ and $A$ decays to $b\bar{b}$}
\label{sec:resultsb}

\begin{figure}
   \begin{picture}(300,570)(0,0)
    \put(50,300){\includegraphics[width=300pt]{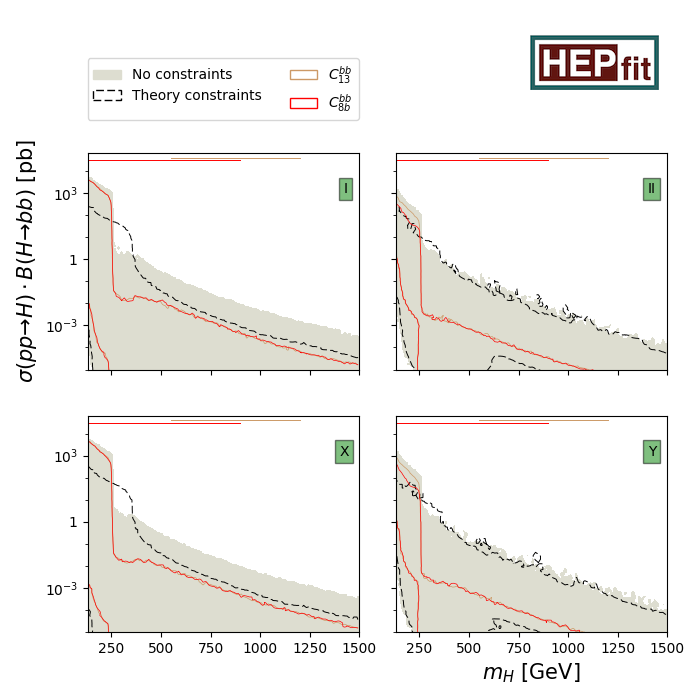}}
    \put(50,0){\includegraphics[width=300pt]{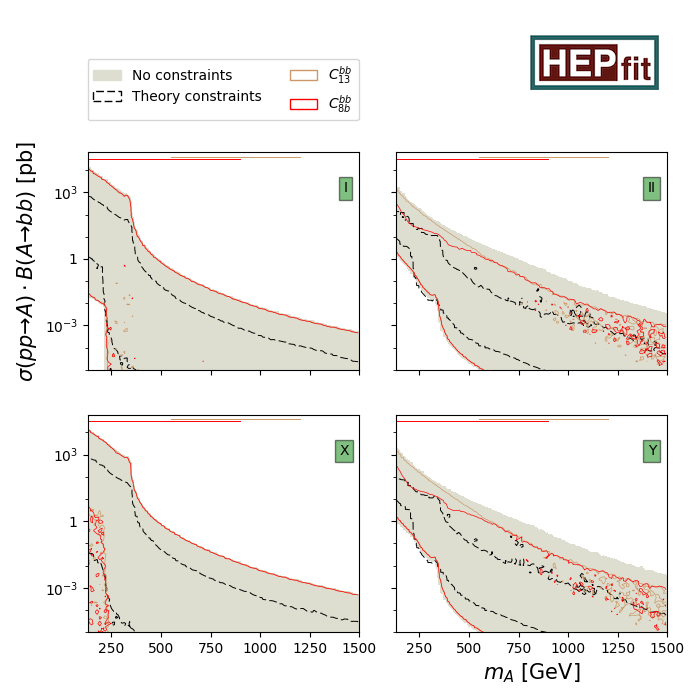}}
   \end{picture}
  \caption{95\% contours of the single searches for $H/A\to b\bar b$ in the \sigbr vs.~$m_{H/A}$ planes for the four 2HDM types (top: $H$, bottom: $A$). For details, see text.}
  \label{fig:Htobb}
\end{figure}

The direct search for $H \to b\bar{b}$ decay does not put any constraint on \sigbr in all four 2HDM types considered in the analysis. In Figure \ref{fig:Htobb} one can see that theoretical constraints provide a suppression of \sigbr by roughly an order of magnitude compared to the fit without any constraint in type I and X. Similar to the previous case, pseudoscalar decaying to $b\bar{b}$ searches do not provide any stronger constraint on \sigbr than the fit with theory constraint alone in all four types. For this search, theory constraints alone restrict \sigbr by at least one order in magnitude with respect to the fit without any constraints in the parameter space analyzed for all the types. This suppression is more dominant in type II and Y compared to the other two cases. In type II and Y, in the regime $m_A \lesssim 600$ GeV the $A \to b\bar{b}$ search from Run 1 suppresses \sigbr compared to the fit without any constraints but it remains sub-dominant or at most of similar strength to the fit with theory constraint alone.

\subsection{$H$ and $A$ decays to $\tau \tau$}
\label{sec:resultstau}

\begin{figure}
   \begin{picture}(300,570)(0,0)
    \put(50,300){\includegraphics[width=300pt]{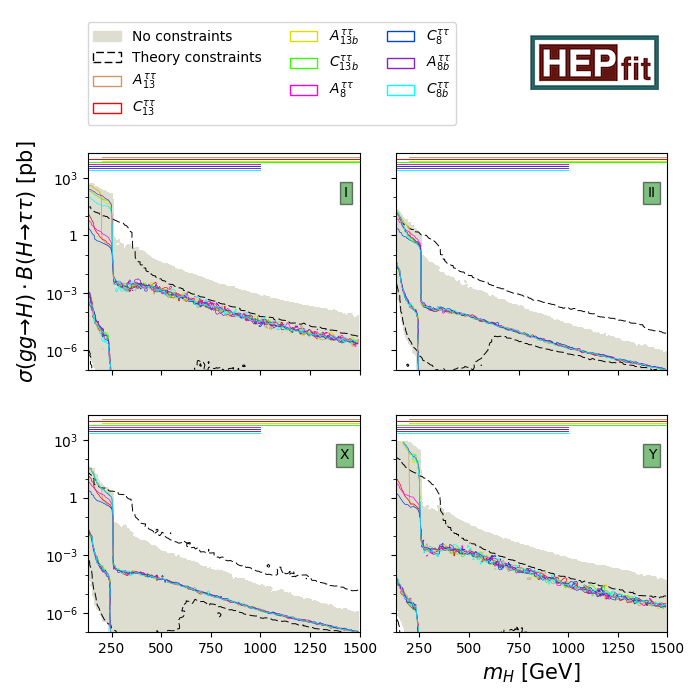}}
    \put(50,0){\includegraphics[width=300pt]{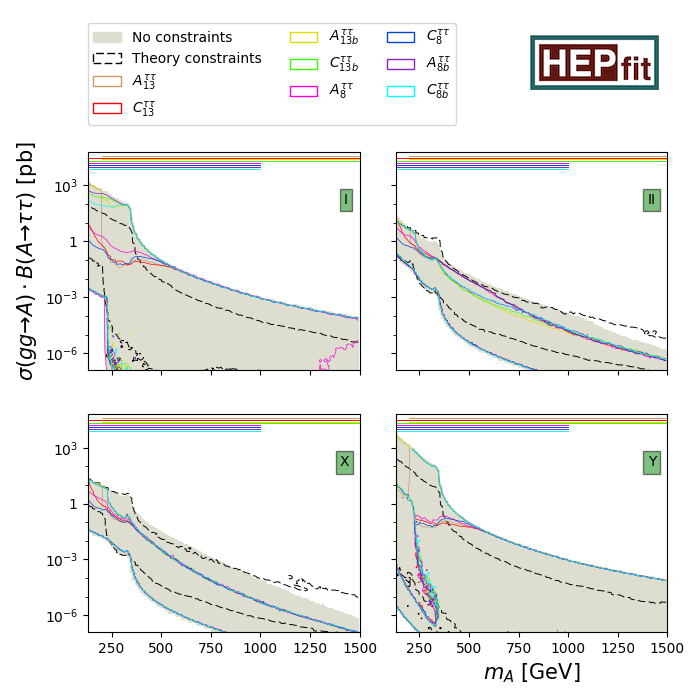}}
   \end{picture}
  \caption{95\% contours of the single searches for $H/A\to \tau \tau$ in the \sigbr vs.~$m_{H/A}$ planes for the four 2HDM types (top: $H$, bottom: $A$). For details, see text.}
  \label{fig:Htotautau}
\end{figure}

In the upper panel of Figure \ref{fig:Htotautau} we show that the $H \to \tau \tau$ searches suppress the \sigbr limit by at least one order of magnitude compared to the fit with theory constraints alone in the regime where the heavy Higgs mass is below 250 GeV. In this regime the strongest constraint comes from Run 1 data and the suppression of \sigbr is more pronounced in the types I and Y. Theory constraints alone restrict \sigbr by roughly an order of magnitude compared to the fit without any constraints for type I and Y, whereas in type II and X, theory constraints raise the lower limit of \sigbr for $m_H > 500$ GeV.
This can be understood as a sensitivity of the fit to fine-tuned scenarios with extreme $\tan\beta$ values, which are disfavoured in a fit to the theoretical bounds.

From the fit without any constraints in the lower panel of Figure \ref{fig:Htotautau} we see that the predicted ranges of \sigbr for the pseudoscalar decaying to $\tau\tau$ are quite narrow for type II and X compared to the other two 2HDM types. The theory constraints yield a suppression by at least one order of magnitude for \sigbr compared to the fit without any constraints in the types I and Y whereas for type II and X the theory constraints push up the lower limit on \sigbr by one order of magnitude compared to the fit without any constraints for $m_A > 350$ GeV. The direct search limits for $A \to \tau\tau$ suppress \sigbr by roughly one to two orders of magnitude compared to the fit with theory constraints alone for all the types except for type Y as long as $m_A \lesssim 300$ GeV.
In type Y, the experimental upper limits on \sigbr are stronger than the theoretical ones for pseudoscalar masses between 200 and 400 GeV, in type II even up to 1 TeV. Scenarios with very light $A$ and \sigbr values between $ 10^{-5} \lesssim$ \sigbr $\lesssim 10 $ pb seem also to be excluded at 95\% by the prior. However, since we have no experimental data on this region, this prior dependence is not an issue here.

\subsection{$H$ and $A$ decays to $\gamma \gamma$}
\label{sec:resultsgaga}

\begin{figure}
   \begin{picture}(300,570)(0,0)
    \put(50,300){\includegraphics[width=300pt]{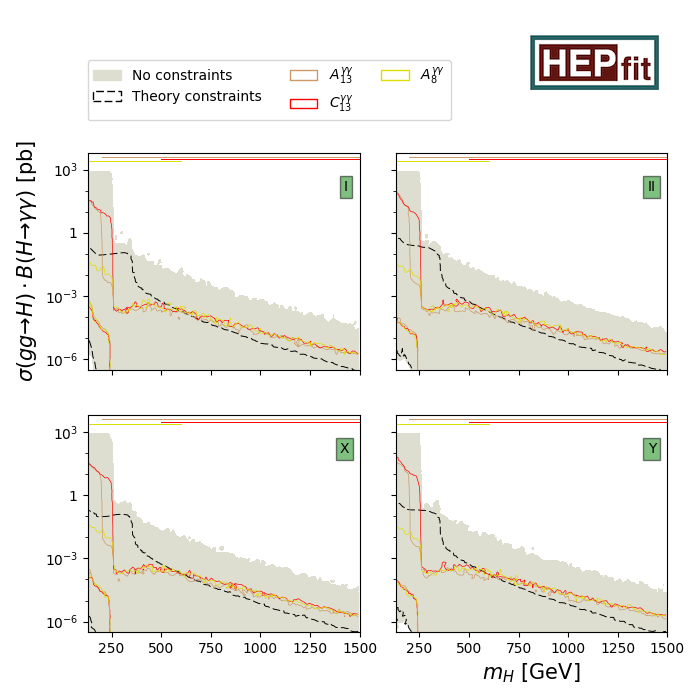}}
    \put(50,0){\includegraphics[width=300pt]{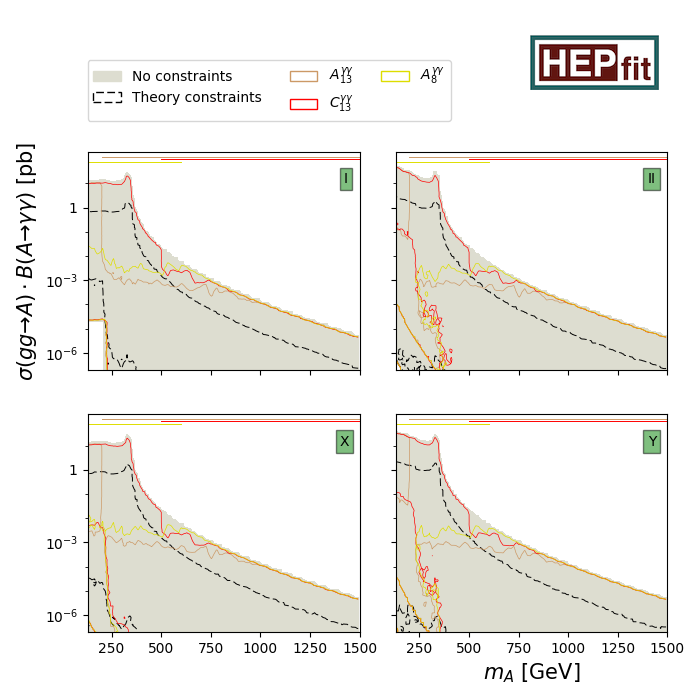}}
   \end{picture}
  \caption{95\% contours of the single searches for $H/A\to \gamma \gamma$ in the \sigbr vs.~$m_{H/A}$ planes for the four 2HDM types (top: $H$, bottom: $A$). For details, see text.}
  \label{fig:Htogaga}
\end{figure}

The theory constraints on $H/A \to \gamma\gamma$ suppress \sigbr by one to three orders of magnitude compared to the fit without any constraints in all four 2HDM types, see Figure \ref{fig:Htogaga}. Direct searches for a heavy CP-even Higgs decaying to two photons constrain \sigbr by roughly one order of magnitude compared the fit with theory constraints for $m_H \lesssim 250$ GeV in all types. The searches in the di-photon decay channel of a pseudoscalar Higgs yield a suppression of \sigbr by one to three orders of magnitude compared to the fit with theory constraints for $m_A \lesssim 600$ GeV for all four types considered. In the types II and Y, we observe again that certain intermediate \sigbr regions for low $m_A$ are disfavoured by the prior.

\subsection{$H$ and $A$ decays to $Z\gamma$}
\label{sec:resultsZga}

\begin{figure}
   \begin{picture}(300,570)(0,0)
    \put(50,300){\includegraphics[width=300pt]{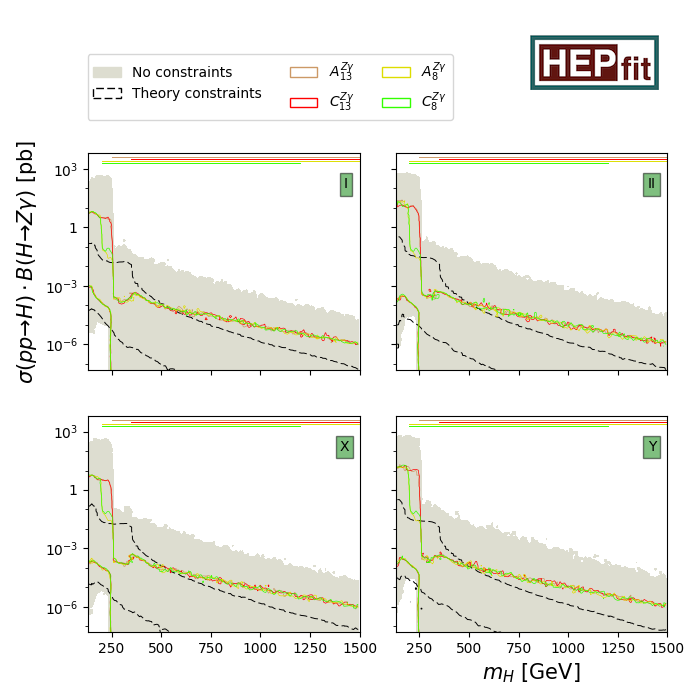}}
    \put(50,0){\includegraphics[width=300pt]{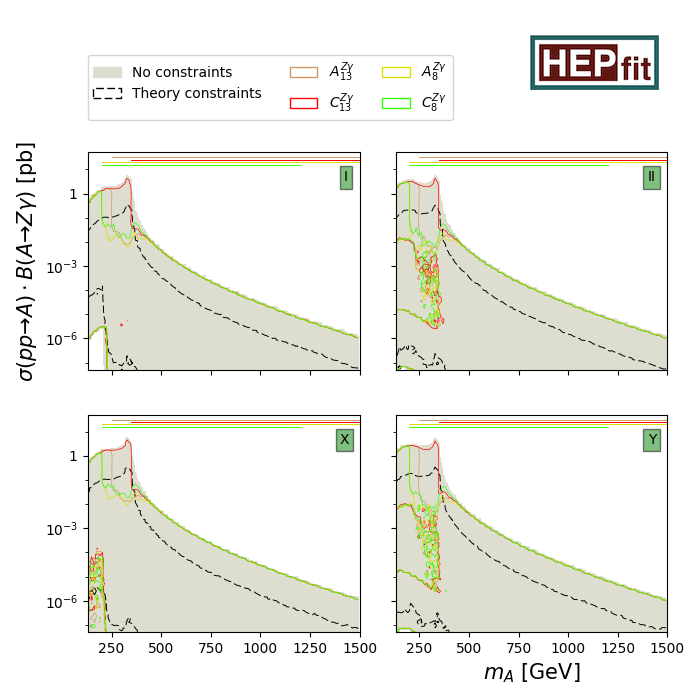}}
   \end{picture}
  \caption{95\% contours of the single searches for $H/A\to Z \gamma$ in the \sigbr vs.~$m_{H/A}$ planes for the four 2HDM types (top: $H$, bottom: $A$). For details, see text.}
  \label{fig:HtoZga}
\end{figure}

We see from the top panel in Figure \ref{fig:HtoZga} that for $H \to Z\gamma$, theory constraints suppress \sigbr by one to four orders of magnitude compared to the fit without any constraints for all types. In all four types, direct search for this channel does not provide any constraint on \sigbr except for a very small window below $m_H \simeq 250$ GeV, but it remains sub-dominant compared to the fit with theory bounds.

Similar to the heavy CP-even Higgs case, theory constraints yield a suppression of \sigbr by one to two orders of magnitude for the decay $A \to Z\gamma$ compared to the fit without any constraints in all four types. Direct searches for this channel provide a suppression of \sigbr by an order of magnitude compared to the fit with theory constraints in the mass window $250 \lesssim m_A \lesssim 350$ GeV in all four types. Again, some parts of the \sigbr region for light $A$ masses are disfavoured by the prior in the types II and Y.

\subsection{$H$ decays to $ZZ$ or $WW$}
\label{sec:resultsVV}

\begin{figure}
   \begin{picture}(300,570)(0,0)
    \put(50,310){\includegraphics[width=300pt]{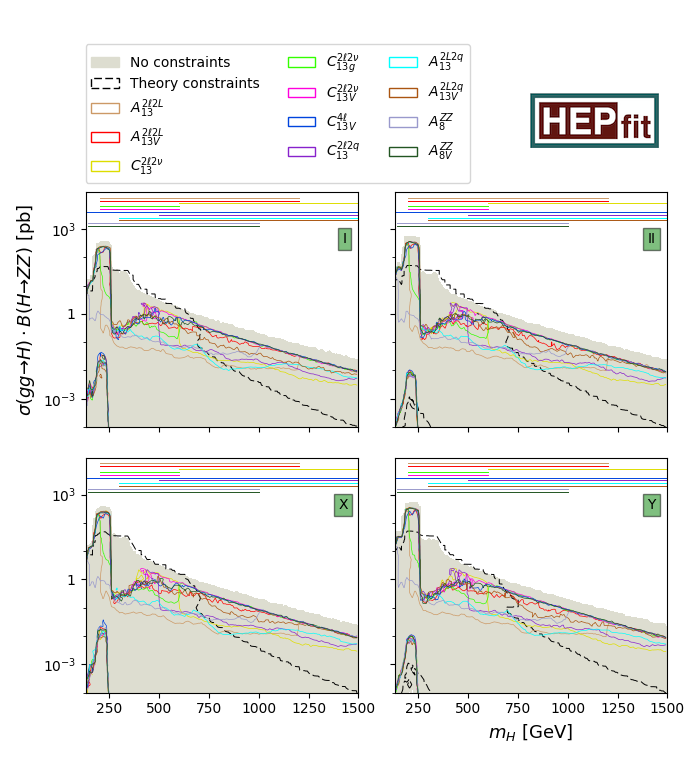}}
    \put(50,-10){\includegraphics[width=300pt]{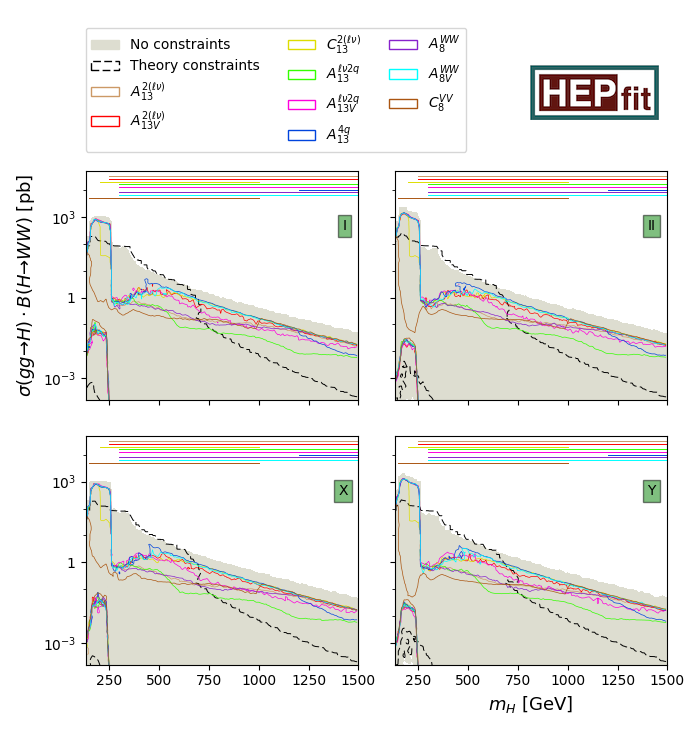}}
   \end{picture}
  \caption{95\% contours of the single searches for $H\to ZZ$ (top) and the remaining $H\to VV$ (bottom) in the \sigbr vs.~$m_H$ planes for the four 2HDM types. For details, see text.}
  \label{fig:HtoVV}
\end{figure}

The heavy Higgs decays to massive gauge bosons can be divided into searches for $ZZ$ and $WW$, but the relative coupling of $H$ to two vector bosons $VV=ZZ,WW$ is universal and type independent. However, the production of the $H$ differs between the types. We show the $H\to ZZ$ channels in the upper panel of Figure \ref{fig:HtoVV} and the searches for $H\to WW$ as well as the combined searches for $H\to VV$ in its lower panel.
The \sigbr are constrained by the theoretical bounds in the decoupling limit, where $m_H>600$ GeV.
The direct LHC searches for this channel yield a strong suppression of \sigbr by one to three orders of magnitude compared to the fit with theory constraint in the mass regime $150 \lesssim m_H \lesssim 800$ GeV ($150 \lesssim m_H \lesssim 700$) for the $ZZ$ ($WW$) channel. For the $ZZ$ searches the $m_H \lesssim 250$ GeV region is constrained by Run 1 data whereas Run 2 data determine the dominant limits for the rest of the region. For the $WW$ searches, Run 1 data dictate the limit until $m_H \simeq 600$ GeV and the high mass range is dominated by Run 2 data. Additionally the $C_8^{VV}$ search for $H\to VV$ severely constrains \sigbr in the $200 \lesssim m_H \lesssim 250$ GeV region for type II and Y.

\subsection{$H$ decays to $hh$}
\label{sec:resultshh}

\begin{figure}
   \begin{picture}(300,570)(0,0)
    \put(50,300){\includegraphics[width=300pt]{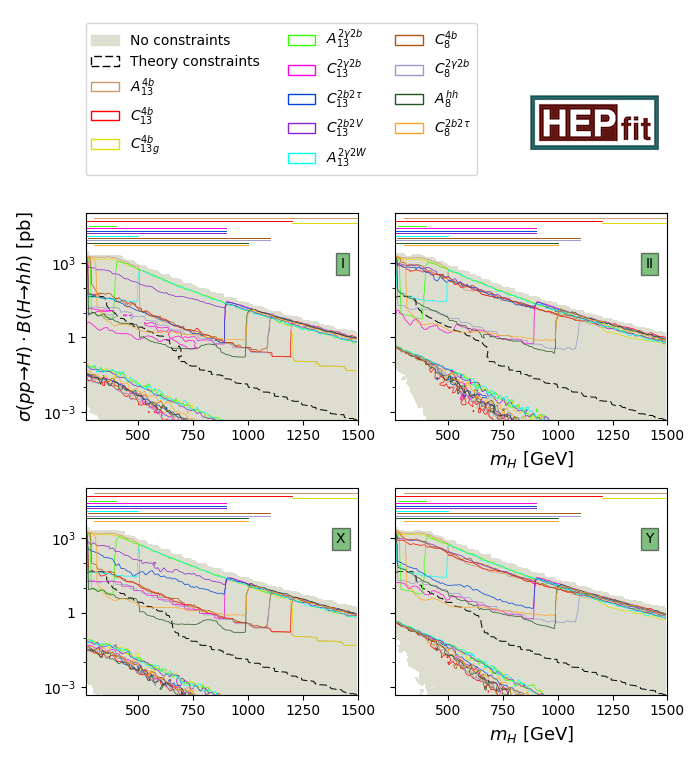}}
    \put(50,-5){\includegraphics[width=300pt]{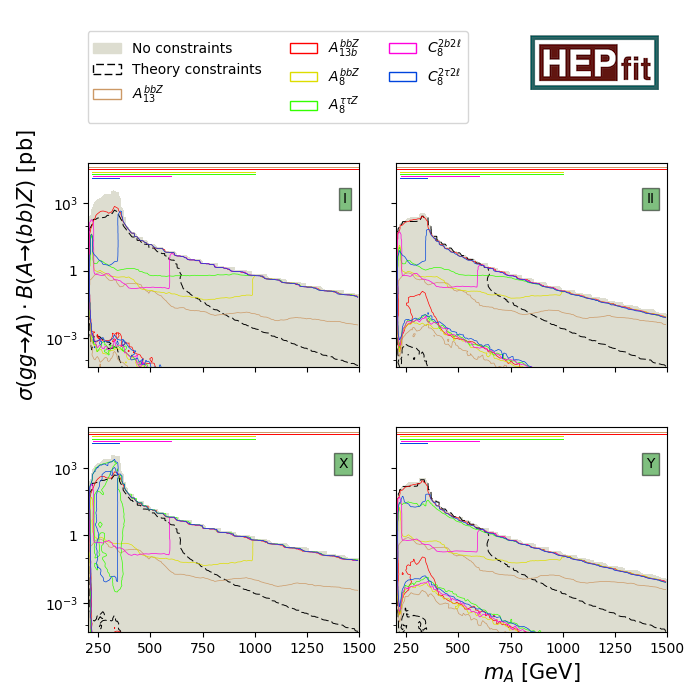}}
   \end{picture}
  \caption{95\% contours of the single searches for $H\to hh$ (top) and $A\to hZ$ (bottom)  in the \sigbr vs.~$m_{H/A}$ planes for the four 2HDM types. For details, see text.}
  \label{fig:Htohh}
\end{figure}

In the upper panel of Figure \ref{fig:Htohh} we show that for the $H$ decaying to two $h$ bosons, theory constraints already yield a strong suppression of \sigbr compared to the fit without any constraints in all four types. Direct searches suppress \sigbr by at most one order of magnitude compared to the fit with the theory constraints in the mass range $200 \lesssim m_H \lesssim 600$ (500) GeV in type I and X (type II and Y). The main constraints stem from Run 1 data and the Run 2 searches for $hh$ resonances decaying to two photons and two bottom quarks.
Although different searches at Run 1 and Run 2 continue to constrain the \sigbr for this channel up to $m_H \simeq 1200$ GeV, they remain sub-dominant to the limit from the fit to the theory constraints.

\subsection{$A$ decays to $hZ$}
\label{sec:resultshZ}

The searches for $A$ decaying into $hZ$ are shown in the \sigbr vs.~$m_A$ planes in the lower panel of Figure \ref{fig:Htohh}; more precisely, they are projected onto the \sigbr of the decay to $(bb)Z$.
As compared to the fit without any constraints, theory constraints effectively are important in the decoupling limit in all types as well as for $m_A \lesssim 300$ GeV in type I and X.
Direct searches for this channel yield a strong suppression (one to three orders of magnitude) of \sigbr compared to the fit with theory constraints for all four types as long as $m_A \lesssim 800$ GeV. In all types, the strongest bounds come from searches in the $A\to hZ\to (bb)Z$ channel; in type X we additionally observe that the $A\to hZ\to (\tau\tau)Z$ limits yield the most important constraints for $m_A$ between 200 GeV and 300 GeV.

\subsection{$H^+$ decays}
\label{sec:resultsHp}

\begin{figure}
   \begin{picture}(300,570)(0,0)
    \put(50,300){\includegraphics[width=300pt]{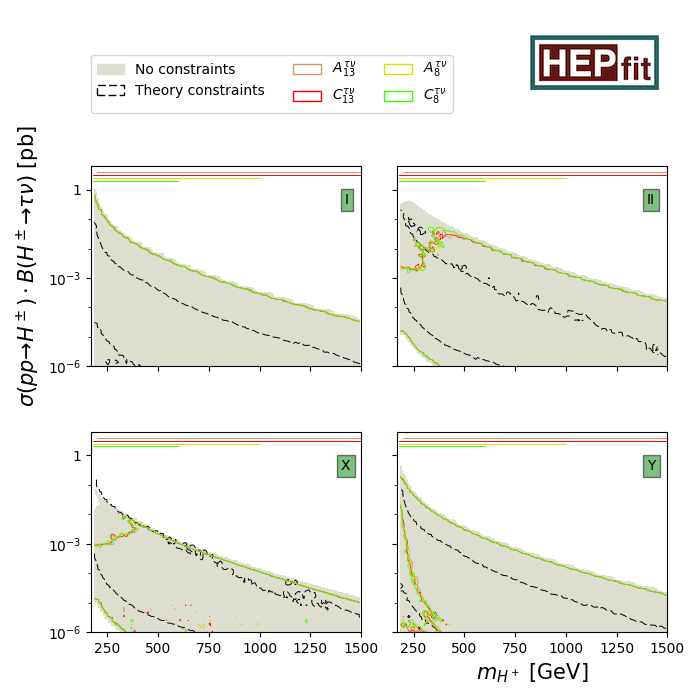}}
    \put(50,0){\includegraphics[width=300pt]{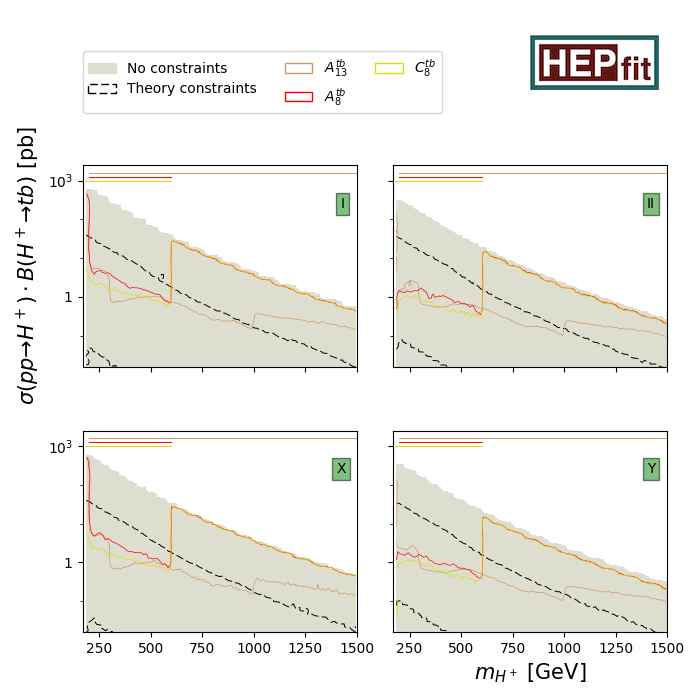}}
   \end{picture}
  \caption{95\% contours of the single searches for $H^+ \to \tau \nu$ (top) and $H^+ \to tb$ (bottom)  in the \sigbr vs.~$m_{H^+}$ planes for the four 2HDM types. For details, see text.}
  \label{fig:Hptotaunu}
\end{figure}

At the present, charged Higgs decaying into $\tau\nu$ does not provide any constraint on \sigbr in the 2HDM's under consideration, see the upper panel of Figure \ref{fig:Hptotaunu}. The theory constraints yield a suppression of \sigbr by roughly one order of magnitude as compared to the fit without any constraint for all types but type X as well as for all types if the charged Higgs decays into $tb$ (lower panel of Figure \ref{fig:Hptotaunu}). In the latter case, the direct searches from Run 1 (Run 2) provide even stronger limits on \sigbr between 180 GeV and 600 GeV (1 TeV), with only small differences between the 2HDM types.

\subsection{All heavy Higgs searches}
\label{sec:resultsheavy}

\begin{figure}
   \begin{picture}(450,525)(0,0)
    \put(0,0){\includegraphics[width=450pt]{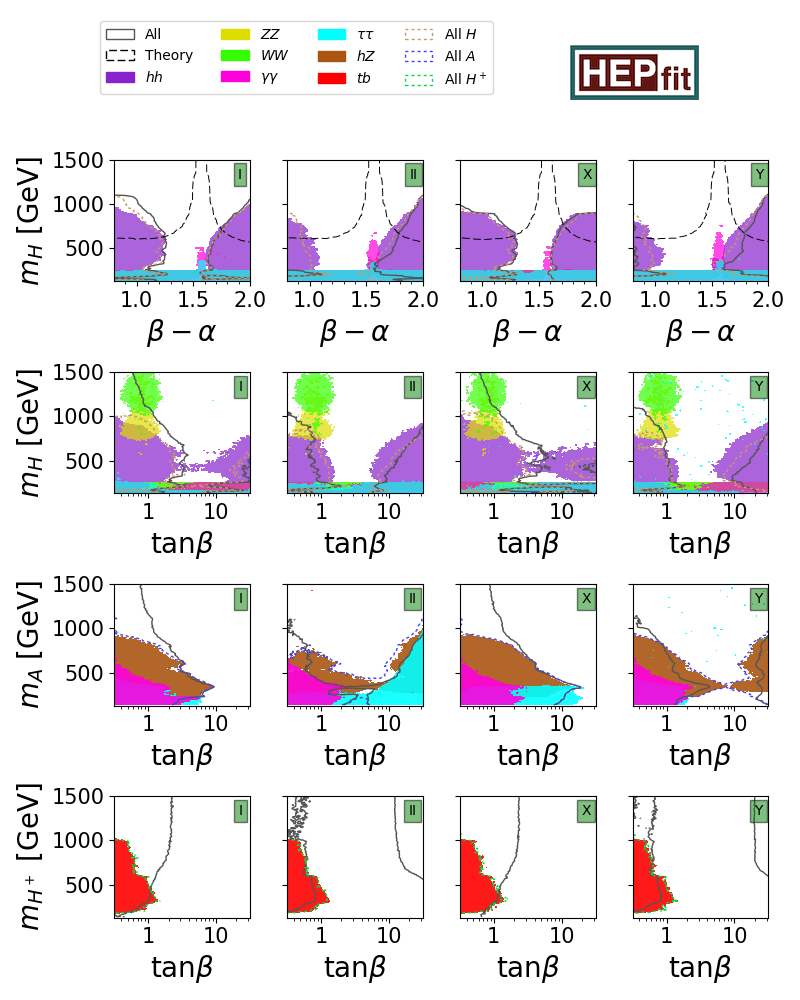}}
   \end{picture}
  \caption{In the 2HDM masses vs.~angles planes we display the regions excluded by all heavy Higgs searches with a probability of 95.4\% by the central area inside the grey solid line. We compare them with the areas excluded by searches in various final states represented by the coloured patches. The areas inside the coloured dashed lines correspond to the exclusion at 95.4\% when all $H$ searches (orange), all $A$ searches (dark blue) and all $H^+$ searches (dark green) are considered. In the first row the limits from theory constraints are shown by black dashed lines.}
  \label{fig:MassAngles}
\end{figure}

In Figure \ref{fig:MassAngles} we show the available parameter space for 2HDM masses and angles from the fit where the heavy Higgs searches are taken into account. The region inside the various coloured patches are disfavoured by the corresponding search category denoted in the legend. The central areas inside the solid grey line mark the 95.4\% allowed regions when all heavy Higgs searches are considered in the fit, including also $C_{8}^{AZ}$ and $C_{8}^{HZ}$. In the panels in the first row the black dashed lines mark the limit from the fit to theory constraints only. The combination of all $H$/$A$/$H^+$ searches is represented by the orange/blue/green dashed contours. The channels described in the previous sub-sections which are not constraining or very weakly constraining in this mass vs.~angles plane are not shown in the figure.

From the first row of Figure \ref{fig:MassAngles} we can see that the region around $\beta-\alpha = \pi/2$ remains unconstrained in all four types of 2HDMs when all the heavy Higgs searches are taken in account. From moderate to high masses, the di-Higgs channels dominate the excluded regions in all four types whereas $H\to\tau\tau$, $H\to\gamma\gamma$ and $H\to VV$ are the most important constraints below the $hh$ threshold. In type I and X, the final exclusion region is mainly dominated by the heavy Higgs to two light Higgs channel, while in type II and Y this is only true if $\beta-\alpha > \pi/2$; for $\beta-\alpha < \pi/2$ the final constraint is weak except for $m_H\approx 250$ GeV.
Although the di-photon searches alone disfavour $m_H \lesssim 600$ GeV for regions near $\beta-\alpha \simeq \pi/2$ this region is allowed when considering all the heavy Higgs searches in the fit. The $H\to VV$ decays only are susceptible to the $\tan\beta\lesssim 1$ region and for $m_H \gtrsim 800$ GeV in all four types; this again is an effect of fine-tuning.
In the displayed $m_H$ vs.~$\tan\beta$ ranges, all the $H$ searches become ineffective for $m_H \gtrsim 1000$ GeV in type II and Y, whereas in type I and X the $\tan\beta \lesssim 1$ regions remain inaccessible even if the heavy Higgs mass is as large as $1500$ GeV. This is an effect of the combination of all heavy Higgs searches, in which scenarios with small $\tan \beta$ are sampled less by the fitter. This feature is not worrisome, because these regions are also suppressed by the flavour observabels as we will show in the next section.

In the pseudoscalar mass vs.~$\tan\beta$ planes we see that the di-photon channel constrains low $\tan\beta$ and $m_A$ up to 600 GeV for all four types. The $A\to hZ$ channel can exclude $\tan\beta$ values up to 10 and $m_A$ almost as heavy as 1000 GeV in all four types. The exclusion also applies for the large $\tan\beta$ regions in the types II and Y. The next most important channel for the pseudoscalar searches in type II is the $A \to \tau\tau$ channel, which efficiently excludes high $\tan\beta$ regions for $m_A$ as large as 1000 GeV. In type X this channel is susceptible to $\tan\beta \lesssim 20$ and $m_A \lesssim 400$ GeV. The exclusion from this channel is weaker for type I and Y. The contour for all pseudoscalar searches is mainly dominated by the $hZ$ channel in type I and Y, and a combination of $hZ$ and $\tau\tau$ in type II and X. Combining all the pseudoscalar Higgs searches the present data constrain certain regions where $m_A \lesssim 1000$ GeV. For type X, the $m_A < 400$ GeV region remains available only if $\tan\beta \gtrsim 10$. In the same mass regions a very narrow range of intermediate $\tan\beta$ remains accessible for type II when all constrains are taken into account. The bound on large pseudoscalar masses is similar to the heavy Higgs case when all constraints are taken into account.

As described in the previous sub-section, the main channel which constrains the charged Higgs mass is $H^{+} \to tb$. The exclusion region of this channel is shown in red in the last row of Figure \ref{fig:MassAngles}. From the figures we see that the present searches for the charged Higgs mass can only constrain the regions with $\tan\beta \lesssim 1$ and $m_{H^+} \lesssim 1000$ GeV in all four types. The inclusion of all the searches in the fit yields stronger constraints in the $m_{H^+}$ vs.~$\tan\beta$ plane than the fit to all charged Higgs searches only. In type II and Y this is due to the $H\to hh\to 4b$ searches, which are particularly sensitive to large $\tan\beta$ and disfavour certain regions featuring $\tan\beta\gtrsim 15$. For type I and X it is more difficult to pinpoint one particular channel for the seeming exclusion of $\tan \beta$ values up to 2 for charged Higgs masses above 1 TeV, but in the next section we see that these bounds can be relaxed once we take into account also other constraints.

\section{Combination of all constraints}
\label{sec:allconstraints}
 
\begin{figure}
   \begin{picture}(450,550)(0,0)
    \put(0,0){\includegraphics[width=450pt]{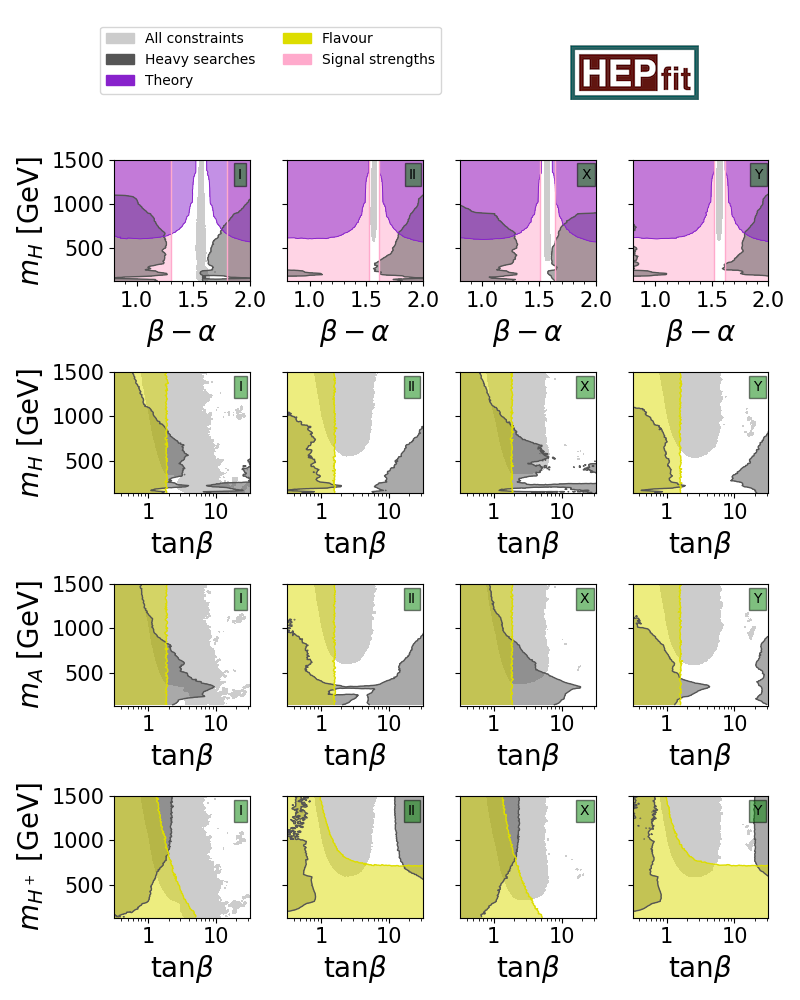}}
   \end{picture}
  \caption{In the 2HDM mass vs.~angles planes we display the regions allowed by all constraints with a probability of 95.4\% in light grey and compare them with the areas excluded by various sets of bounds: The 95.4\% contours of heavy Higgs searches (dark grey), flavour observables (yellow) and $h$ signal strengths (pink) as well as the 99.7\% limits from theory constraints (purple).}
  \label{fig:MassAnglesAll}
\end{figure}

After discussing the individual effects of the $h$ signal strengths and the searches for $H$, $A$ and $H^+$ on the 2HDM, we want to confront these constraints with the other bounds on the parameters. In Figure \ref{fig:MassAnglesAll} we copy the information about all heavy Higgs searches from Figure \ref{fig:MassAngles} in the mass vs.~angle planes, and add the bounds from signal strengths and theoretical constraints to the $m_H$ vs.~$\beta-\alpha$ planes and the impact of the flavour observables to the planes with $\tan \beta$. Finally, the global fit to all constraints is represented by the grey regions.

The signal strength bounds do not depend on the masses of the heavy Higgs particles; their limits on $\beta-\alpha$ differ for each type, see Section \ref{sec:signalstrengths}. The theory conditions force the 2HDM's into the alignment limit for $m_H>600$ GeV, decoupling the heavy Higgs particle from physics around the electroweak scale. The type dependence of this effect is negligible as it only enters via sub-leading Yukawa terms in the beta function parts of the NLO unitarity conditions. Besides the obvious consequences for certain constellations of $m_H$ and $\beta-\alpha$, the combination of the signal strengths with the theory constraints also disfavours large values of $\tan \beta$. This is because extreme values for the latter result in a destabilization of the unitarity conditions \cite{Cacchio:2016qyh}. Specific combinations of the 2HDM angles can still fulfil the theoretical constraints, but these solutions are highly fine-tuned and thus have a low posterior probability.
The flavour constraints have been discussed many times in the literature; the summary is that in all types the $B_s$ mass difference sets lower limits on $\tan \beta$ (at least for masses within the reach of the LHC), while the branching ratio of $b\to s \gamma$ processes enforces $m_{H^+}\gtrsim 580$ GeV at 95\% C.L.~in type II and Y \cite{Misiak:2017bgg}. In combination with theory and electroweak precision bounds, these limits on the charged Higgs mass can be translated to lower limits on the neutral masses. (The individual impact of $STU$ will be explained below as it is not visible in these two-dimensional projections of the parameter space.)
 
The combination of all constraints is more intricate than the naive superposition of all individual bounds. First of all, we should mention that it also depends on the prior we choose for the masses: while the direct experimental observables depend on the masses of the heavy Higgs bosons, the theoretical bounds and the loop-induced effects are only sensitive to the mass squares. Since we want to combine both, we need to decide whether we want to use a flat prior for the masses or the mass squares. A detailed discussion can be found in Appendix \ref{sec:appendixA}. The contours we show here are a superposition of a fit with flat mass priors and a fit with flat mass square priors in order to be as conservative as possible.
In the $m_H$ vs.~$\beta-\alpha$ planes of type I and X, the combination of signal strengths and theory only leave a very small strip around the alignment limit of $\beta-\alpha=\pi/2$. Heavy Higgs searches additionally exclude $m_H<380$ GeV in type X. The reason for this are not only the $H\to \tau\tau$ searches as observed in Figure \ref{fig:MassAngles}, as they are similar to the bounds in type I, but mainly it is an interplay of the strong bounds of the $A$ searches for low $\tan\beta$ and the above-mentioned exclusion of large $\tan\beta$ due to signal strengths and theory, which together disallow $m_A<400$ GeV. This bound translates to a limit on $m_H$ using the unitarity and electroweak precision constraints, since these bounds delimit the mass splittings, see below. Also in the type II and Y planes we see a lower limit of $m_H>550$ GeV, which in this case derives from the lower bound on $m_{H^+}$ from the $b\to s \gamma$ measurements and the fact that the mass difference $m_H-m_{H^+}$ cannot be too large.
The absolute maximal deviation of $\beta-\alpha$ from $\pi/2$ is 0.03 in type I and 0.02 in the types II, X and Y. (This corresponds to $1-\sin(\beta-\alpha)<5\cdot 10^{-4}$ and $<2\cdot 10^{-4}$, respectively.)
In the $m_H$ vs.~$\tan \beta$ planes one can see that the fine-tuning for large $\tan\beta$ scenarios disfavours these regions and pushes the allowed contours towards smaller values of $\tan\beta$. Only in type I and for $m_H<350$ GeV, the heavy Higgs searches have a visible impact on this plane, excluding $\tan\beta\lesssim2.5$. More or less the same holds for $m_A$ vs.~$\tan \beta$, where in type I $\tan\beta<3$ is excluded by direct $A$ searches if $m_A<350$ GeV. We already mentioned above that in type X, the interplay between fine-tuning and $A$ searches sets a lower limit of 400 GeV on $m_A$.
Having a look at the $m_{H^+}$ vs.~$\tan \beta$ planes, one can see the lower bounds on the charged Higgs mass in type II and Y from $b\to s \gamma$, which we quantify to be 600 GeV in our fit.
Also here, we observe that large $\tan \beta$ values are disfavoured and the posterior regions are shifted towards small $\tan \beta$.

\begin{figure}
   \begin{picture}(450,520)(0,0)
    \put(0,0){\includegraphics[width=410pt]{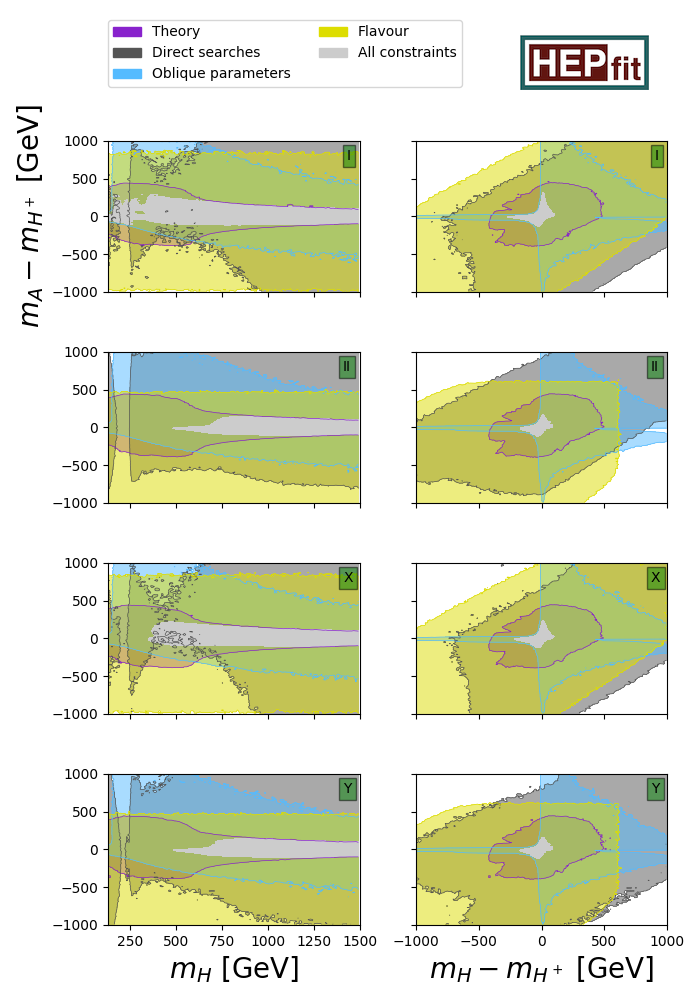}}
   \end{picture}
  \caption{In the $m_A-m_{H^+}$ vs.~$m_H$ (left panels) and $m_A-m_{H^+}$ vs.~$m_H-m_{H^+}$ (right panels) planes we show the allowed regions by various sets of constraints: The heavy Higgs searches, the oblique parameters and the flavour observables determine the 95.4\% allowed contours in dark grey, light blue and yellow, respectively. For the theoretical constraints, the 99.7\% regions are given by the purple shaded areas. We superimpose the 95.4\% probability combination from the global fit to all observables in light grey.}
  \label{fig:MassDifferences}
\end{figure}

The fit only to electroweak precision data does not exclude any region in the two-dimensional mass vs.~angle projections in Figure \ref{fig:MassAnglesAll}. What it does constrain are the mass differences between $H$, $A$ and $H^+$. That is why in Figure \ref{fig:MassDifferences} we show the difference between the pseudoscalar and charged Higgs mass, once depending on the $H$ mass (left column) and once against the $m_H-m_{H^+}$ difference.
In the $m_A-m_{H^+}$ vs.~$m_H$ planes the dominant constraints come from the theory bounds, at least in the decoupling limit $m_H>600$ GeV. The $STU$ pseudo-observables are stronger if $m_H<600$ GeV and $m_{H^+}>m_A$. In type I, they yield a lower bound on $m_A-m_{H^+}$ in the global fit for $m_H<250$. If we look at the $m_A-m_{H^+}$ vs.~$m_H-m_{H^+}$ planes, we observe that here the oblique parameters are the strongest constraint on $m_A-m_{H^+}$ if the charged Higgs mass is larger than $m_H$.
Combining all constraints and marginalizing over all other parameters, we obtain the following ranges for the mass differences allowed with a probability of 95\%:\\
\begin{table}[htb]
\centering
\begin{tabular}{| l | c c c |}
\hline
&$m_H-m_A$ [GeV]&$m_H-m_{H^+}$ [GeV]&$m_A-m_{H^+}$ [GeV]\\
\hline
Type I & [-202;82] & [-158;75] & [-104;183]\\
Type II & [-127;69] & [-120;62] & [-92;110]\\
Type X & [-168;79] & [-134;70] & [-98;155]\\
Type Y & [-130;70] & [-130;60] & [-90;110]\\
\hline
\end{tabular}
\end{table}

\noindent We can thus exclude the decays $H\to H^+H^-$, $H\to AA$, $H\to H^+W^-$ as well as $H\to AZ$ in all four types with a probability of 95\%.

For all heavy Higgs search limits, we implicitly assumed the narrow width approximation. In a simultaneous fit to all constraints except for these direct search limits we find that with a probability of 95\% the decay widths of $H$, $A$ and $H^+$ never exceed $5.5\%$ of the mass of the particle in the types II and Y. For masses below 1 TeV the maximal decay widths are less than $3.5$\% in these two types.
In type I and X and for $\phi=H,H^+$ the fits yield $\Gamma_\phi/m_\phi<3.5$\% ($<5$\%) if $m_\phi<1$ TeV ($<1.5$ TeV). Only the ratio $\Gamma_A/m_A$ can reach $7$\% for $m_A\approx 550$ GeV, but in the decoupling limit $m_A>600$ GeV similar bounds apply as for $H$ and $H^+$. We would like to stress here that all these $95$\% limits are \textit{maximally} allowed values and that in a typical 2HDM scenario the widths are significantly smaller.
Therefore we conclude that the narrow width approximation is a reasonable choice for 2HDM scenarios.

Finally, addressing the last variable of our chosen parametrization, we also observe limits on the soft $\mathbb{Z}_2$ breaking parameter $m_{12}^2$. The upper limits strongly depend on the maximally allowed physical Higgs masses and are around $(1$ TeV$)^2$ in all types. Due to the lower mass limits on the physical Higgs particles in the types II, X and Y, we also observe that $m_{12}^2$ is limited from below, the respective minimal values being (280 GeV)$^2$, (170 GeV)$^2$ and (240 GeV)$^2$. Only in type I an unbroken $\mathbb{Z}_2$ symmetry is still compatible with all constraints.

The run time for the global fits to all constraints with 240 million iterations was $550\pm 130$ CPU hours.

\section{Conclusions}
\label{sec:conclusions}
 
In all four 2HDM types with a softly broken $\mathbb{Z}_2$ symmetry we have presented global fits to the most recent data.

Focussing on the latest measurements from LHC, we have showed explicitly how the individual signal strengths affect the leading order $h$ couplings at tree-level and at one-loop level. Combining all information about the signal strengths, we find that the quantity $|\beta-\alpha-\pi/2|$ cannot exceed 0.26, 0.055, 0.069 and 0.056 in the types I, II, X and Y. The one-loop couplings of the $h$ to gluons and photons cannot differ by more than 20\% and 30\%, respectively, relative to their SM values.

In order to systematically discuss the searches for $H$, $A$ and $H^+$, we have categorized them according to their decay products and have compared the exclusion strength of the single available ATLAS and CMS analyses on the production cross section times branching ratio, depending on the masses. We have then combined all decay categories and have showed their impact on the 2HDM masses and mixing angles. For $m_H$ below 1 TeV we observe strong bounds on ``extreme'' values for the angles, that is if $\beta-\alpha$ is very different from the alignment limit $\pi/2$ or if $\tan \beta$ is smaller than 1 or larger than 10. The exact limits depend on the model type and $m_H$. Also the LHC searches for pseudoscalars severely constrain the 2HDM parameters: For $m_A<1$ TeV, the lower limits on $\tan \beta$ reach values of around 10 in the types I and X. In the types II and Y, these limits are weaker, but there are also mass dependent upper limits. The bounds from charged Higgs searches are less constraining in comparison; nevertheless, they also start to be stronger than the indirect constraints in the regions with low $m_{H^+}$ and low $\tan \beta$.

Finally, we have confronted the LHC $h$ signal strengths and heavy Higgs searches with all other relevant indirect constraints from theory and experiment. In detail, we have showed how stability and unitarity constraints and $B$ physics observables set mass dependent limits on the 2HDM angles and on the differences between the heavy Higgs masses, while electroweak precision data only affect the latter. We have compared all different sets of constraints and have showed the results in the mass vs.~angle planes as well as in the $m_A-m_{H^+}$ vs.~$m_H(-m_{H^+})$ planes for all four types of $\mathbb{Z}_2$ symmetric 2HDM's together with the simultaneous fit to all constraints. In this global fit we find the following 95\% probability limits on the 2HDM parameters marginalizing over all other parameters:
$|\beta-\alpha-\pi/2|$ cannot be larger than $0.03$ in type I and $0.02$ in the other types. 
In type II and Y, $m_H>700$ GeV, $m_A>750$ GeV, $m_{H^+}>740$ GeV and $m_{12}^2>(240$ GeV)$^2$, while we observe lower mass limits of $m_H>450$ GeV, $m_A>500$ GeV, $m_{H^+}>460$ GeV and $m_{12}^2>(170$ GeV)$^2$ for type X. For the latter, it is the first time that a statistically significant lower limit on the massive parameters has been observed in a global fit for the analyzed mass ranges. Also, if we discard particularly fine-tuned scenarios, only the following ranges for $\tan \beta$ are  allowed for masses below $1.6$ TeV: [0.93; 10.5] in type I, [0.93; 5.0] in type II, [0.93; 5.2] in type X and [0.91; 5.6] in type Y.
However, the upper limits are no strict bounds and have to be taken with a grain of salt.
Moreover, we can put type dependent upper limits of order of 100 GeV on the differences between $m_H$, $m_A$ and $m_{H^+}$, and thus kinematically exclude all decays of $H$ or $A$ into another heavy Higgs particle. As a consequence, the decay widths of $H$ and $H^+$ cannot exceed $5.5\%$ of their mass in all types, at least as long as we consider masses below $1.5$ TeV. While in the types II and Y we see a similar limit for the $A$ decay width, it can amount to up to 7\% of $m_A$ in the types I and X.

\acknowledgments
We thank Enrico Franco, Ayan Paul, Maurizio Pierini and Luca Silvestrini for useful discussions. The fits were run on the Roma Tre Cluster. We are grateful for the availability of these resources and especially want to thank Antonio Budano for the support. We also thank Jose Miguel No and Ken Mimasu for giving us a hand with the implementation of the $H\to AZ$ and $A\to HZ$ search limits. This work was supported by the European Research Council under the European Union's Seventh Framework Programme (FP/2007-2013) / ERC Grant Agreement n.~279972, by the Indo-French Center for Promotion of Advanced Research/CEFIPRA (Project no.~5404-2) and by the Spanish Government and ERDF funds from the European Commission (Grants No.~FPA2014-53631-C2-1-P and SEV-2014-0398).

\appendix
\section{Prior dependence of the massive parameters}
\label{sec:appendixA}

\begin{figure}
   \begin{picture}(450,560)(0,10)
    \put(0,0){\includegraphics[width=450pt]{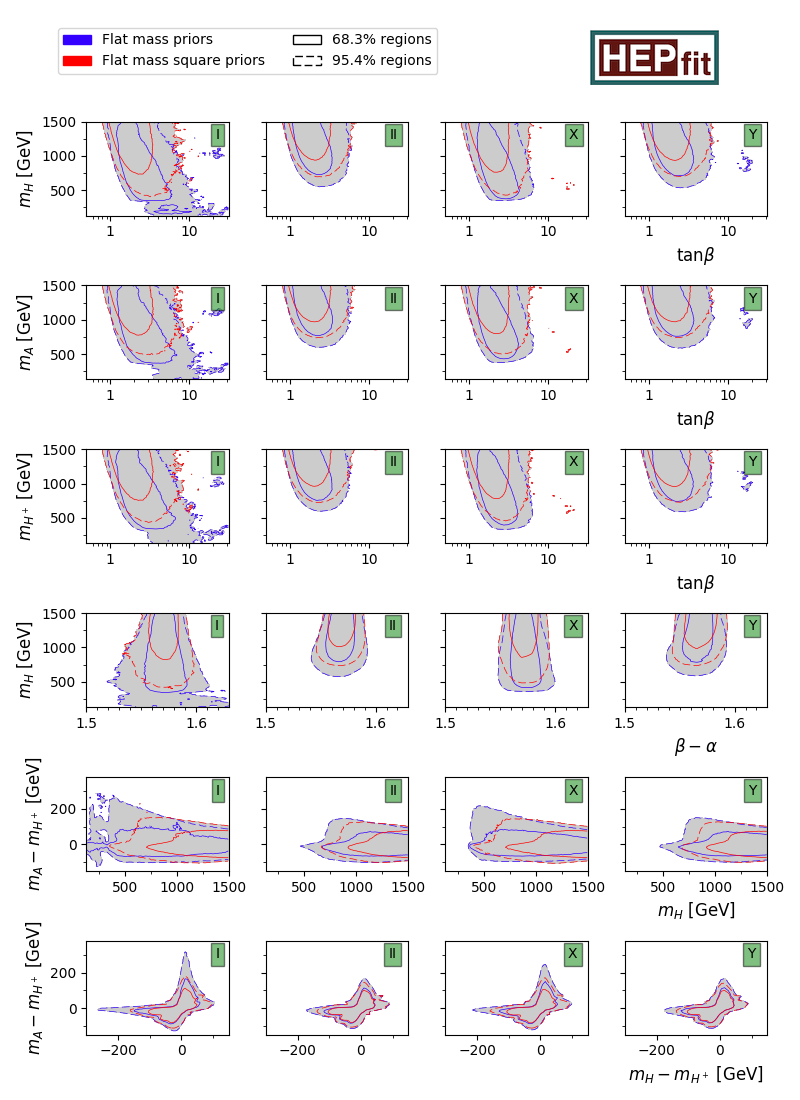}}
   \end{picture}
  \caption{The mass prior dependence of the fit with all constraints is shown in the $m_{H/A/H^+}$ vs.~$\tan \beta$ planes, in the $m_H$ vs.~$\beta-\alpha$ planes (corresponding to Figure \ref{fig:MassAnglesAll}) and in the $m_A-m_{H^+}$ vs.~$m_H$ and $m_A-m_{H^+}$ vs.~$m_H-m_{H^+}$ planes (corresponding to Figure \ref{fig:MassDifferences}), from top to bottom.}
  \label{fig:allcomparison}
\end{figure}

In this appendix we discuss the prior dependence of our analysis when \textit{all} the constraints have been taken into consideration. In Figure \ref{fig:allcomparison} we compare the allowed parameter space in the mass versus angles planes and the mass difference versus mass (difference) planes from the fit with flat mass priors (blue solid and dashed curves) with the fit with flat mass square priors (red solid and dashed curves). At a first glance one can see that the fit with flat mass square priors prefers high mass regions and raises the lower limit on the masses by $\mathcal{O}(100)$ GeV compared to the fit with flat mass priors.

It is well known that Bayesian statistics do not provide a unique rule to determine the prior distribution and in general the posterior distribution is a prior dependent quantity. A thumb rule would be to choose a flat prior for the parameter on which the observables depend linearly. For example, if an observable quadratically depends on a particle mass, one would choose a flat mass square prior. Unfortunately, in the 2HDM the theoretical and indirect experimental constraints depend on the mass squares, whereas the direct experimental observables dependent on the masses. Not having sufficiently constraining data makes assigning the mass priors in the fit with all constraints is a delicate task. This would not be problematic if the observables were measured with a high precision; in fact, we can see that in the types II and Y the difference between the two priors are considerably smaller as there are strong lower mass limits.
In order to be as conservative and prior independent as possible, we decided to combine the 95.4\% regions for both priors: The light grey contours for the fits with all constraints in Figures \ref{fig:MassAnglesAll}, \ref{fig:MassDifferences} and \ref{fig:allcomparison} are obtained by superimposing a fit with flat mass priors and a fit with flat mass square priors. The corresponding numerical results mentioned in Section \ref{sec:allconstraints} are based on the more conservative fit; for instance, the limits for masses and mass differences were extracted from the fit with flat mass priors, while the upper limits on the decay widths were larger in the fits using flat mass square priors.

%
\bibliographystyle{JHEP}
\bibliography{thdmupdate2017}

\end{document}